\documentstyle[11pt,aaspp4]{article}

\input{psfig}
\psdraft
\psdraftbox
\psfull

\def\U{U_{300}}
\def\B{B_{450}}
\def\V{V_{606}}
\def\I{I_{814}}
\def\UB{(U_{300}-B_{450})}
\def\BV{(B_{450}-V_{606})}
\def\VI{(V_{606}-I_{814})}
\def\BVVI{(B_{450}-V_{606})/(V_{606}-I_{814})}
\def\UBBV{(U_{300}-B_{450})/(B_{450}-V_{606})}
\def\Ly{Ly$\alpha$ }

\begin{document}

\begin{center}
To appear on the February 1999 issue of the {\it The Astronomical Journal}
\end{center}

\vskip 72pt

\title{\Large\bf{Quasar Candidates in the Hubble Deep Field$^1$}}

\vskip 12pt

\author{Alberto Conti, Julia D. Kennefick${^2}$, Paul Martini, Patrick S.
Osmer} \affil{Department of Astronomy, The Ohio State University, \\
Columbus, OH, 43210, USA.\\
conti, martini, osmer@astronomy.ohio-state.edu, julia@astro.ox.ac.uk}

\vskip 24pt

\altaffiltext{1}{Based on
observations with the NASA/ESA Hubble Space Telescope obtained at the
Space Telescope Science Institute, which is operated by Association of
Universities for Research in Astronomy, Incorporated, under NASA
contract NAS5-26555.}

\altaffiltext{2}{Currently at the NAPL, Keble Road, Oxford (UK), OX1 3RH} 

\vskip 60pt

\centerline
{\bf Abstract}

The Hubble Deep Field gives us an unprecedented view of our 
universe and an opportunity to study a wide range of questions 
in galaxy evolution and cosmology.
Here we focus on the search for unresolved faint quasars and AGN 
in the crude combine images using a multicolor imaging 
analysis that has proven very successful in recent years.
Quasar selection was
carried out both in multicolor space and in
``profile space''. The latter is the multi-parameter space formed by
the radial profiles of the objects in the different images. By combining the
dither frames available for each filter, we were able to obtain well
sampled radial profiles of the objects and measure their deviation 
from that of a stellar source. This procedure efficiently helps to
overcome the problems caused by the WPFC2
undersampling. 
Furthermore, to identify areas of multicolor 
space where quasars might be expected, we generated
synthetic quasar spectra in the range  
$1.0<z<5.5$ including effects of intrinsic emission lines
and absorption by \Ly forest and Lyman-limit 
systems, and computed expected quasar colors. 
We also developed routines to determine the completeness of our data
to point sources in the observed filters. The data are $90 \%$ complete 
at 26.2$^m$, 28.0$^m$, 27.8$^m$, 26.8$^m$ in the F300W, 
F450W, F606W and F814W filters respectively.\\ We find
41 compact objects
in the HDF: 1 pointlike object with colors consistent with quasars but
inconsistent with that of ordinary stars, 7 pointlike objects with
colors consistent 
with quasars or stars; 18 stars; and 15 slightly resolved objects,
12 of which have colors consistent with quasars or stars.  The latter
objects could be resolved active galactic nuclei or regions of intense
star formation.   We estimate the upper limit of unresolved and slightly
resolved quasars/AGNs with $\V < 27.0$ and $z < 3.5$ 
to be 20 objects (16,200 per deg$^2$). We independently recovered a
resolved, star-forming galaxy with a spectroscopically confirmed
redshift of 3.368 and 5 spectroscopically confirmed stars. Both provide
confirmation for the validity of our color and morphological modeling.\\
We make a detailed comparison with previous studies of the HDF. We find
good agreement among different authors on the number of stars in the HDF
and the lack of quasar candidates with $z>3.5$. We find more quasar candidates
than previous work because of our more extensive modeling and use of
the color information. Spectroscopic observations of the candidates
are needed to determine which, if any, are quasars or AGNS.


\clearpage
\section{Introduction}

The study of the high redshift universe has seen a dramatic
development over the last two years with the discovery of a large population
of star-forming galaxies at redshifts $z \sim 2-4$ (\cite{abr96};
\cite{ell97}; \cite{cle96}; \cite{mao97}; \cite{els96}; \cite{zep97};
\cite{ste96}; \cite{low96}; \cite{mob96}; \cite{lan97};
\cite{phi97}; \cite{mad96}; \cite{guz97}; \cite{pet97}). The Hubble Deep 
Field (HDF, \cite{wil96}) has played a key role in this development,
since it provides not only the deepest images for multi-color
selection of ``Lyman break'' galaxies but also morphological and size
information for high redshift galaxies (\cite{bou97}; {\cite{ber98}).
As we go to fainter and fainter magnitudes,
morphological information for individual objects becomes harder to extract
(\cite{abr96}) and the best approach to understand these objects is
through a multi-dimensional analysis of
the populations of objects as a whole. 
From this perspective, objects detected in the HDF can be regarded
as the components of a multi-dimensional parameter space, whose axes include
at a minimum the apparent magnitude, three colors, and an angular size.
Larger and brighter galaxies  are also interesting in their own right
because the morphologies of these relatively high-redshift objects can
be compared in detail to those of present day galaxies; these sorts of
studies are being carried out by other authors, such as the
members of the Medium Deep Survey team (\cite{roc96}; \cite{mut94}).

Acquired during $\sim 150$ orbits near the north galactic
pole in December 1995 and imaged in four filters: F300W ($\U$),
F450W ($\B$), F606W ($\V$) and F814W ($\I$), the data for the HDF resulted 
from 35 hours of exposure time in $\B$, $\V$  and $\I$  and 45 in $\U$.
The unique feature of the HDF images is their ability
to resolve extremely faint extended sources and to distinguish them from
point sources down to an unprecedented magnitude (see Section
\ref{sec:cplt}). Several authors have also used this features to put
constraints on the nature of the dark matter (Flynn, Gould and Bahcall
1996 hereinafter FGB; \cite{col97}; \cite{kaw96}) and on galaxy and
structure formation 
in the universe (\cite{met96}; \cite{bau96}; \cite{col96};
\cite{saw97}; \cite{van96}). It is natural to make use of this
intrinsic property to look for faint quasars in the HDF. 

Quasars are believed to be the visible manifestation of the
accretion of matter onto supermassive black holes and, being among the most
distant and luminous objects in the universe, hold a primary role as
cosmological probes. Due to their stellar-like nature,
quasars cannot be distinguished from stars in single images of the sky, as
galaxies can. However, multi-color selection techniques have proven
very successful in recent years in selecting quasar candidates at high redshifts
(\cite{war94}; \cite{ken95}; 
\cite{hal96} and references therein). At lower redshifts ($z<2.2$), quasars can be selected
in the $\U$ and $\B$ images by adopting the standard
UV-excess technique. As \Ly moves into the $\V$ filter, quasars
at $z>3.5$ can be selected on the basis of the $\B$-$\V$ color, and as
\Ly moves into the $\I$ filter and continuum absorption by \Ly becomes
stronger, the $\V$-$\I$ color will serve as a powerful
selector of quasars at $z>4.8$. In addition, by making use of the
excellent angular resolution of HST, we can add a morphological
selection criterion to our color selection criteria. This will enable
us to select those objects 
whose morphological {\em and} color characteristics are consistent
with them being quasars or active galactic nuclei (AGN). 

The main purpose of this paper is to find spatially unresolved quasar
candidates which have colors consistent with quasars in the range
$1<z<5.5$.
Section~\ref{sec:det} explains the trade-offs involved in
the choice of the lower resolution crude combine images as opposed to the
higher resolution drizzled images. As observed below, we chose to use the lower
resolution crude combine images for object detection together with
the raw dither positions for PSF modeling because they improved results on
scales smaller than 1 pixel.

Section~\ref{sec:cplt} describes the steps involved in computing the
detection completeness of the HDF to point sources. We did not
simulate artificial fields with HDF-like properties, instead we used
the original crude combine images and carefully added artificial
point sources in a wide range of magnitudes. A model PSF was created
for each chip and filter from a subsample of the stars found in the HDF by FGB.
The number of detected point sources as a function of magnitude
provides stringent limits on our detection efficiency at faint
magnitudes and will be used to constrain object classification.

Direct object classification from the crude combine images is not
possible due to the undersampling of the WFPC2. In
Section~\ref{sec:class} we describe the steps used to overcome this
problem and obtain spatial information on objects within a 1 pixel 
radius from their center.
This allowed us to produce a list of objects with characteristics
typical of stellar-like sources. In Section~\ref{sec:color}, by
generating synthetic quasar spectra in the redshift range $1<z<5.5$ and
convolving them with the HDF filter curves, we determine what colors
to expect for quasar candidates. Then we combine the spatial
and color information to produce a list of candidate quasars in
Section~\ref{sec:cand-sel}. In Section~\ref{sec:discussion} we present
an estimate of the upper limit of quasar candidates and discuss the results.
We compare our results with previous work in Section~\ref{sec:comp} and
consider future searches for AGNs in Section~\ref{sec:future}.

\section{Image Selection and Source Detection \label{sec:det}}

In order to reduce the photometric errors due to flatfielding
uncertainties, the HDF team made the decision, early in the planning
stage, to carry out the HDF observations at roughly 9 different
pointing positions per filter, spanning a range of $2.6\arcsec$. 
This procedure called ``dithering'' has the secondary purpose of
trying to recover some of the information lost due to the undersampling of the
WFPC2 detectors. This is achieved by allowing sources to be sampled in
different portions of a pixel at each dither position. 

The HDF data were reduced two times. Version 1 accounts for processing up
to January 1996. Version 2 was released February 1996 and contains
a new set of images obtained with a new technique
for the linear combination of images known formally as variable-pixel
linear reconstruction and informally as ``drizzling'' (\cite{fru96}).
The drizzled images produced have a new pixel scale 2.5 times smaller
than the original $0.1\arcsec$ typical of the WFPC2. While higher
resolution is an attractive feature, the interpolation scheme used
alters the radial profiles of the sources in the images and
introduces correlated noise among neighboring pixels. 
These effects result in a widening of the PSF from as little as $5\%$ for
Poisson noise free images, to as 
much as $15\%$ for real HDF images. Correlated noise results in an
underestimated pixel-to-pixel rms noise by a factor of 1.9 (\cite{wil96}).

We thus decided to use the Version 1 crude combine images for object
detection and photometry. These are 
images at different dither positions shifted (with integer pixel
shifts) and combined with a median filter to produce
a median image at the original pixel scale with no correction for geometric
distortion, but with a minimal amount of interpolation. 
Even though individual images show curvature of a few percent in the
background  within 200 pixels of the edge of the chip,
these variation are reduced to less than $1\%$ when multiple
frames are stacked together (\cite{wil96}). 

The use of the median in the image combination may lead
to systematic errors in the photometry if the images are not 
perfectly registered. The HDF team checked the registration of the
individual images by comparing the positions of several bright
sources. They found that the images at each dither position were 
registered well within the errors of such a comparison. Two 
unexpectedly large (0.8\arcsec) shifts have been treated as separate
dither positions (\cite{wil96}).
We are therefore confident that accurate photometry can be performed
on the crude combine images. 
However, to reliably distinguish stars from
galaxies at faint magnitudes, we need to be able to sample the radial
profile within the inner central pixel (FGB). To achieve
this accuracy, we revert to the use of the raw dither images in all
four filters. 
Section~\ref{sec:class} will describe why such a combined approach is
crucial to a reliable object classification.

We used the FOCAS analysis package (\cite{jar79}; \cite{val82}) for object 
detection run under the SKICAT catalog management software
(\cite{wei95}) to detect objects in the HDF crude combine images.
To prepare the frames for object detection, we trimmed them
of bias regions and registered them to the nearest pixel in all four filters.  
To optimize our detection algorithm we tested a range of convolution
kernels (a series of Gaussian kernels and the FOCAS built-in one) for
data smoothing 
in order to 
find the one most suitable for the objects in the HDF. In general,
spurious detections are introduced by either a poor noise estimate
around the object or by object sizes that are not well matched by the
detection kernel. The Gaussian filters ranged in FWHM from
that of the point-spread-function of the HDF ($\sim 1.2$ pixels) to 
10 times this width. The Gaussian filter with twice the FWHM of the HDF 
outperformed the other Gaussian filters. The FOCAS built-in filter, however, 
performed the best overall by maximizing the number of detected objects
while minimizing spurious detections.  

Consequently, we ran FOCAS on all images and, after the value of each pixel was
convolved with the FOCAS built-in detection kernel, we required each
detected object to have a minimum of $6$ contiguous pixels, or 0.09
square arcseconds, above a detection threshold of 2.5 times the
local sky rms. FOCAS then examines these sources for sub-components.
This process, however, can break large bright unmistakably
recognizable spiral galaxies into many subcomponents which are not
independent objects.
In order {\it not} to include these objects in
our catalog, we manually merged them back into their parent.
A total of 24 objects in 9 galaxies were re-merged.

The newly formed $\U,\B,\V,\I$ catalogs were
matched using the pixel coordinates. 
Aperture photometry was performed for each object in the $\U,\B,\V,\I$
catalogs using the IRAF\footnote{IRAF is distributed by the National
Optical Astronomy Observatories, which are operated by the Association
of Universities for Research in Astronomy, Inc., under cooperative
agreement with the National Science Foundation.} PHOT task with an aperture of
2 WFPC2 pixel radii or $0.2\arcsec$. 
The output from PHOT was then used with the HDF Vega zeropoints and the
exposure times in each filter to derive magnitudes for all objects.
Photometric errors as a function of magnitude returned by PHOT are
shown in Figure~\ref{fig:magerr}.
Color-color diagrams for $\UB$ vs $\BV$ and $\BV$ vs $\VI$ of all the
objects in the HDF with a $5\sigma$ detection in the 4 filters are shown in
Figures~\ref{fig:ubbv} and \ref{fig:bvvi} respectively.
The figures show clearly that most of the objects have $-0.2<\BV<1.4$
and $0<\VI<1.5$. The lack of a very red compact population is evident
and not entirely new. Current findings from medium-deep surveys
emphasize the presence of the rising fraction of faint systems with
irregular morphologies suggestive of rapid stellar evolutionary
processes (\cite{gla95}; \cite{dri95}).

\section{Completeness \label{sec:cplt}}

A fundamental step to achieve a useful catalog of the objects
in the HDF is to understand how often stellar objects failed to be
detected as a function of magnitude in each filter
and in each separate WFPC2 chip, i.e. the completeness limits of the
dataset to point sources.
To determine the completeness of the HDF, we adopted the
straightforward approach of adding artificial 
point sources to the HDF frames using the IRAF/ARTDATA
task over a wide range of magnitudes. Adding
artificial point sources to the actual HDF frames does not require the
generation of artificial images that mimic the observed properties of
the HDF and has the further advantage of matching the noise properties of the
data.

In performing any completeness test, particular attention must be paid to 
the PSF of the added sources. This should resemble as closely as
possible that of the actual data, including its noise properties. 
We decided to construct a PSF template using the sample of stars found
in the HDF by FGB. We selected unsaturated stars
brighter than $\V \sim 25$ in each chip.
We compared each candidate star radial profile with a PSF generated by
TinyTim (\cite{kri93}), a sophisticated PSF simulation software that
takes into account chip and optics distortions to generate a model
PSF. To further improve our PSF templates, we derived 
radial profiles from a sample of stars extracted from the  
Hubble Space Telescope PSF Library. The agreement among these three
model PSFs was excellent. However PSF variations among different
chips prompted us to generate PSF templates for each of the 3 chips in
each filter rather than to produce a global HDF PSF. This approach,
while adding a further layer of complexity to the analysis, will 
turn out to be particularly useful during the morphological
classification phase. A template PSF used in the
completeness test for Chip 4 in the $\I$ filter is shown in
Figure~\ref{fig:psf-crude}. The figure shows the distance in pixels
from the 
center of the template against the fraction of light falling on that
pixel. The flux has been normalized to the total flux within a 4 pixel
radius aperture. The WFPC2 undersampling is evident. 

It should be noted that TinyTim implicitly assumes that the
source is always centered 
on the central pixel. This is clearly visible in Figure~\ref{fig:psf-crude}
where all the filled circles, representing the TinyTim model, are
exactly positioned at fixed distances from the central pixel. This
artifact is solely introduced by the modeling of TinyTim. Real
sources will be randomly positioned with respect to the central pixel
and hence produce a PSF that samples a wider range of distances
from the center. We will in fact show in Section~\ref{sec:class} how the
actual position of the source can be used to produce a well sampled
PSF well within the innermost central pixel, thereby allowing an
accurate morphological separation between stars and galaxies.

To measure the completeness of the HDF to point sources, we added 10
template stars to each chip in each filter ten times
for a total of 100 stars per magnitude bin.
We then applied the same detection procedure adopted for the original HDF
frames, using FOCAS, and recorded the number of added point sources that
we recovered as a function of magnitude.
In the range of input magnitudes $23<m<32$ we
observed a slow linear decline before the sharp cutoff at the detection
threshold. This change in behavior occurs at the $\sim 90\%$
completeness level. In Figure~\ref{fig:cplt} we show the percentage of
added template point sources that were recovered as a function of
magnitude in each chip. Error bars represent a $1\sigma$ uncertainty
around the mean value. 

As clearly shown in Figure~\ref{fig:cplt}, there are small differences
among the filters in different chips. At the $90\%$ completeness
threshold the $\B$ and $\V$ data have, within the errors, the same
behavior and reach depths that differ on average by only $0.1$
magnitudes. However, while at $90\%$ completeness the $\B$ image is deeper
than the $\V$ image, there is an inversion at the $50\%$ level where the
$\V$ image becomes deeper. 
In Table~1 we list the $50\%$ and
$90\%$ completeness for all chips in each image. 

\section{Compact Source Selection \label{sec:class}}
Bahcall et al. (1994) have shown how radial profiles can be
effectively used to  
separate stars from galaxies. However, to distinguish stellar
from extended profiles we need at least 5 times more photons than
required for a simple detection (FGB). We will therefore restrict our
classification to all the objects with a $5\sigma$ detection in $\B$,
$\V$ and $\I$ ($\U$ being optional), and $\sim 1$ magnitude brighter than
the $90\%$ completeness limits 
for each filter. The resulting total number of objects detected in
($\U$) $\B$ $\V$ $\I$ is 1709.
Hence our search for point-like
sources will not include objects fainter than $\U \sim 25.2^m$, $\B
\sim 27.0^m$, $\V \sim27.0^m$, $\I \sim 26.0^m$.

The steps undertaken to produce a reliable classification are
described in detail below and are summarized as follows:
\begin{enumerate}

\item the objects in the crude combine images
were shifted according to all the available dither positions in each
filter (from a minimum of 8 to a maximum of 11);

\item all objects were identified separately in each dither and
their radial profiles computed;

\item all dither profiles were then combined together to produce a
densely sampled profile of each object on sub-pixel scales;

\item each profile was then compared to that of the PSF and its
departure from the PSF measured;

\item simulations were used to predict the expected departure as
a function of magnitude.
\end{enumerate}

A fundamental step in constructing the radial profiles is determining
the location of
the center of the object to better than $0.1$ pixels by making use of
the symmetry of the object about its center. To overcome the 
poor PSF sampling of the WFPC2, we used a technique that combines the
raw dither frames to obtain a well sampled radial profile in each
filter of all our objects. This task is accomplished by producing a
shifted catalog of objects (using the image shifts made available by
the HDF team) for each dither position, leaving the single
images untouched and therefore avoiding any flux interpolation among
aligned frames. Each catalog is then used as an input to the
IRAF/RADPROF task that computes the radial profiles and accurately
re-centers them.
We used 9, 8, 11 and 8 separate dither positions in $\U$, $\B$, $\V$ and
$\I$ respectively. The different number of dithers used reflects the
exclusion of frames affected by the characteristic X-pattern due to 
scattered light from the Earth (\cite{wil96}). For each object, the
single dither 
profiles obtained are then combined together to produce a well
sampled radial profile that will in turn be used for
classification (Figure~\ref{fig:psf-dither}).

The few unsaturated
stars were used to construct PSF templates upon which to base
classification. Each PSF template was in turn compared with radial
profiles from a large number of stars that appear in the WFPC2
exposures at low galactic latitude. Again, the agreement was
excellent.
The advantage of using single dither profiles to overcome the WFPC2
undersampling is evident in Figure~\ref{fig:psf-dither} when compared to the
crude combine profile in Figure~\ref{fig:psf-crude}. Since the radial profile
is sampled in many adjacent pixels at different distances from the
image center, adding dithers together produces a densely
sampled profile well within the innermost pixel.
Figure~\ref{fig:psf-dither} shows the radial profile of a $21.2^m$ magnitude
star. Each filled circle represents a pixel
in one of the 11 dithers used in constructing the profile. The flux has been
normalized to the total flux within 4 pixels. 

To better model the PSF, we fit a Moffat profile (\cite{mof69}) to the few
unsaturated stars in each chip and in each filter of the following form:
\begin{equation}
I(r) = I_0 \left(1+\left(\frac{r}{\alpha}\right)^2\right)^{-\beta}
\end{equation}
where $I_0$ is the central intensity and $\alpha$ is a measure of the
width of the profile and can be easily
related to its FWHM.
The choice of a Moffat profile as
opposed to, for example, a Gaussian, was motivated by the parameter
$\beta$, which  allows for a better modeling of the data in the wings
of the distribution where most of the points reside.
To further constrain our fit, we also paid much care to the way the weights
were assigned to the single 
points. We decided to weight more ``signal'' rather than ``noise'' by
assigning larger weights in the fit to the innermost points (those
within 0.6 pixels of the center). After normalizing
the profile to the total flux within the innermost 4 pixels, we chose
weights that are proportional to the Poisson noise at each point of
the profile.
For stellar sources like the one in
Figure~\ref{fig:psf-dither}, the normalized profile contains more
than 90\% of the total light within the innermost 1.4 pixels.
Furthermore, by assigning small weights to points at large distances from
the center of the object, we force our fitting algorithm to focus on
the amount ($I_0$) and concentration ($\alpha$) of the signal rather
than allowing it to find low $\chi^2$ solutions driven by the
concentration of points at distances greater than 2 pixels. 
During classification, as we
will later show, this approach will essentially reduce the available
degrees of freedom in our fit from 3 to 2, thereby simplifying our
further analysis. 

Once all the Moffat parameters were found and the 12 PSF templates created
(one for each chip in each filter), we fit a Moffat profile to all
the remaining objects in our catalog, setting the value of the
parameter $\beta$ to that of
the appropriate (chip and filter) PSF. The parameter $\beta$ controls
the ``rate of decay'' of the tail of the Moffat distribution. By
fixing $\beta$ to its appropriate value, we constrain our fitting
algorithm to minimize $\chi^2$ for a given tail dependency.
This procedure returns $I_0$
and $\alpha$ (or the FWHM of the profile),
which are now directly comparable to those of the template PSFs.
Among these parameters, $I_0$ is not only the most stable and physically
meaningful but, most importantly, is the most sensitive for
discriminating between stars and extended objects.
Figure~\ref{fig:galaxy} shows one such fit for a galaxy. 
For comparison, the PSF template is also shown. It is 
clear that $I_0$ is indeed a well suited parameter
for star/galaxy separation. Candidate selection, however, requires a
further step: the assessment of the scatter in $I_0$ as a
function of magnitude for stellar objects.

For the selection of the stellar candidates, we decided to simulate
stars of different magnitudes and compute their Moffat parameters as
was
done for real HDF objects. In this particular case, however, we
need to simulate a ``dither star'' of a given magnitude (i.e. a single
star viewed at separate dither positions) and then combine its dither
radial profiles to form a densely sampled profile in exactly the same way
as was done for the real data. This approach is again quite robust
since the artificial stars added to the single dithers match perfectly
the noise properties of that particular dither, filter and chip.

Template stars were created using the IRAF/ARTDATA task where the
PSF parameters for a Moffat profile were taken from the relevant chip and
filter. We added 10 stars per magnitude bin in the range $23<m<29$,
computed their radial profiles, fitted a Moffat profile and recovered
$I_0$ and $\alpha$. Figure~\ref{fig:Vclass} shows
the typical behavior with magnitude of the central intensity parameter
$I_0$ for all objects in the $\V$ filter. 
The dashed lines represents the $95\%$
confidence interval (given by our simulations) within which stellar
candidates lie. As the template star magnitude approaches the
completeness limit of the data, the errors increase past the
point of reliable classification. This limit, in magnitude, is roughly
$\sim 0.75^m$ brighter than the $90\%$ completeness limit in each
filter, with the exception of the $\U$ filter where the limit is $\sim 1^m$
brighter than the 90\% completeness limit due to the low quantum
efficiency of the $\U$ detector which is 
virtually noise limited, as opposed to the other filters which are nearly
sky-noise limited (\cite{wil96}).

Using the central intensity parameter $I_0$ and its
variation as a function of magnitude, we can now select
compact objects in each filter. We
find 57, 36, 32 and 15 objects in $\U$, $\B$, $\V$, $\I$ respectively.
However, not all of these 
objects have detections in all 4 filters (mainly because of missing
$\U$ detections). For candidate selection we will require a candidate
to be selected as compact in at least one of the 4 filters. The number
of independent objects meeting this criterion is 41. These objects are
listed in Table~2 along with their magnitudes and
coordinates both on the sky and on the drizzled images.

For classification purposes 
we will select objects classified as compact in at least one of the
three redder filters: $\B$ or $\V$ in the $\UBBV$ plane and $\V$ or $\I$
in the $\BVVI$ plane respectively. All 41 objects in Table 2 meet this
requirements. This first stage of our analysis efficiently extracts all
the compact sources in the dataset down to the 90\% 
completeness limit. One confirmation of the effectiveness of our
selection algorithm is that we recover all 17 stellar-like objects found
by eye by FGB (see Table~5). While most of these objects
are relatively bright 
sources whose profiles are unmistakably classified as stars (both by
an eye search by FGB and our algorithm), as we approach the
completeness limit of the dataset, galaxy contamination becomes important
and a clear classification is not possible on morphological grounds
only (see Figure~\ref{fig:Vclass}). Color information becomes
therefore an essential tool to help us distinguish among compact galaxies,
stars and quasars.

Before proceeding in our analysis, however, we realized that
our morphological parameter space allows for an additional assessment of 
the ``degree of compactness'' of each object independent of color.
In fact, the quantity $I_0$ does not in itself
provide for a measure of the deviation of a given object's profile from that
of the PSF but acts more like a minimum threshold as a function of
magnitude for selecting compact objects.
While it may seem that combining $I_0$ with the FWHM or $\alpha$ would be a
better choice to assess this deviation, our tests on simulated stars
show that a more stable measure of such a deviation as a function of
magnitude is obtained by taking the ratio of the peak to total flux
within 4 pixels and comparing it to that of known stars in the HDF. This new
parameter can be formally defined as:
\begin{equation}
\Delta = 1 - R_{obj}/R_{psf}
\end{equation}
where $R_{obj} =  I_0/I_{tot}$ is the ratio of the peak to total flux
for each of our 41 objects and $R_{psf}$ that of the PSF. 
The parameter $\Delta$ was used for classification only in the three
redder filters because higher noise in the $\U$ images made them
not useful for this purpose.
As done for $I_0$, we computed $\Delta$ for our simulated stars and
obtained the 
range of variation of $\Delta$ as a function of magnitude and filter.
With this new quantity in hand, we can now separate stellar from
compact objects in a more objective fashion and hence try to reduce
galaxy contamination for those objects close to the limiting
magnitudes for classification. We will require objects to be within
the $3\sigma$ confidence intervals of the PSF in at least 2 of the 3 redder
filters in order for them to be classified as quasars. Our confidence
intervals as a function of magnitude are tabulated in Table~3.

As an immediate application, we measured $\Delta$
for all 17 objects found by FGB (see Table~5). As
mentioned before, 
most of these objects are quite bright and are easily classified as
stars based on the appearance of their radial profiles, the value of
their $I_0$ parameter and the value of $\Delta$ (see
Table~2). We then decided to re-examine only the 6
faintest objects in the FGB sample (all of which have $\V>26.5$) in
order to assess whether any of them had 
been erroneously classified as stars. These 6 objects, together with
the additional 24 that we classified as compact, constitute our final
sample. This sample of objects will be 
analyzed in detail in the following section, where we will use color
information for the objects along with
simulated quasar and stellar colors to divide all the
objects in the sample into three groups: stars, quasars and compact
galaxies. The latter refers to those objects that do indeed appear compact,
according their $I_0$ value, but either have a significant deviation from the
stellar profile (indication of the presence of an extended
component) or do not exhibit colors consistent with stars or quasars.

\section{Color Selection \label{sec:color}}
Having morphologically selected compact source candidates, we return now
to the use of multicolor space to investigate the presence of objects
whose colors are consistent with them being quasars. This method has been
widely used, with much  success, in the past (\cite{new97},
\cite{miy97}; \cite{hal96}; \cite{ken95}; \cite{sie95}; \cite{war94};
\cite{war91})  and its application is
quite straightforward.
To determine what colors to expect for
quasar candidates, we first created synthetic
quasar spectra at redshifts $1 < z < 5.5$ in steps of $\Delta z = 0.1$.
In generating these spectra, we assumed
a power law continuum of the form $f_{\nu} \propto \nu^\alpha$
and Gaussian profile emission lines of FWHM=5000 km s$^{-1}$. 
Rest frame equivalent widths were taken from Wilkes et al. (1986) for
Ly$\alpha$, N V, 
Si IV, C IV, and C III, and from Warren et al. (1991) for O VI.  

Wilkes et al. (1986) give the rest frame equivalent widths (EW) of
Ly$\alpha$ of a typical quasar 
as 65\AA. We have considered three line strength cases by varying the 
strength of the Ly$\alpha$ line, but keeping the EW ratios fixed:
EW(Ly$\alpha$) = 32.5, 65.0, and 130.0. This range of EW includes 90\%
of the objects in the Wilkes et al. sample (95\% if we include
two more objects that fall outside the range by only 1\AA). 
We have also allowed for a variation in the continuum slope by
creating spectra with $\alpha$ = -0.50, -0.75, and -1.00 for each line
strength and redshift. 
Another important factor in computing the expected colors of quasars is 
the continuum drop blueward of Ly$\alpha$ due to intervening absorption.
We have used models of the absorption due to the Ly$\alpha$ forest,
Lyman-limit systems and Lyman continuum absorption as a function
of redshift.  For full details, see Kennefick at al. (1995) and
references therein.
For each redshift, line strength, and continuum slope considered, we
generated five different quasar spectra with different realizations
of the absorbing systems for a total of 1845 synthetic spectra.

We convolved the synthetic spectra with the HDF filter curves to obtain
simulated quasar colors. Photometric magnitude errors were added to the
simulated colors according to the distribution shown in Figure
\ref{fig:magerr}; we assumed a Gaussian  
distribution of errors at a given magnitude down to $\V= 29.4$,
corresponding to a $\sim 3\sigma$ detection. We show in Figures 9 and 10
the loci that contain 99\% of the simulated quasars. We will use these
loci to define the regions in color-color space for quasar
candidates. We consider this to be the most suitable approach due to the
lack of a quasar database in the four WFPC2 filters used for the
HDF. Kennefick et 
al. (1997) have shown for the DMS ground based survey that an identical
modeling approach is in good agreement with the observed quasar
colors. 

The colors of the morphologically selected compact sources in the HDF
can now be directly compared to those of model quasars in Figures 9 and
10. The observed points are denoted by circles. 
For objects not detected in $\U$, we adopted a 
3$\sigma$ limiting
magnitude value of 26.9, according to the results shown in
Fig. 1.  This limit is based on photometric measurements
and error estimates of all the detected objects in the HDF.
In this case, the limit corresponds to a flux value 3$\sigma$
above zero. 

It is worth
mentioning that since we did not require objects to be detected in $\U$,
the presence or lack of a $\U$ detection does not limit our subsequent
color analysis or the limiting magnitude we give below in \S 7. In fact,
Figure 9 shows that there are no objects with color upper limits that
would overlap our quasar locus for low 
redshift objects. As for the $\BVVI$ color-color plane, the $\U$
information was not used to select high-z candidates.
We also compare the colors of our objects to those of typical stars.
The stellar points were
derived from the Bruzual-Persson-Gunn-Stryker (BPGS) spectrophotometric
atlas available through the STSDAS/SYNPHOT package under IRAF. This atlas is
an extension of the Gunn-Stryker optical atlas into both the UV and the
infrared (\cite{gun83}). The IR 
data are from Strecker et al. (1979) and the IR and optical data were tied
together by the $V-K$ colors. The transformation from standard Johnson/Cousin
passband colors to HST colors was obtained using SYNPHOT by
convolving the spectrum of each of the stars in the BPGS 
spectrophotometric atlas with the HST filter and chip response curves.

\section{Candidate Selection \label{sec:cand-sel}}

Now we are able to use the index of compactness, $\Delta$, 
and the color information in Table 2 to classify each object
as a quasar candidate, star, or compact galaxy\footnote{A detailed
description of the color and morphological properties of the compact
sources in the HDF is available from the first author.}.  We will consider
objects with $\Delta < 3 \sigma$ as starlike and those
with $\Delta > 3 \sigma$ as compact galaxies.  We use the color information
to place the objects in one of three categories: 1) colors on or within
the quasar locus and away from the stellar locus; 2) 
colors consistent with either quasar or stars; or 3) colors not
consistent with quasars.  Objects in categories 1 and 2 are
divided according to whether their colors are consistent with
low-redshift quasars (lz - ultraviolet bright, see Figure 9) or 
high-redshift quasars 
(hz - see Figure 10).
Objects in category 3 are classified
as stars or galaxies according to their $\Delta$ parameter.

The results of this classification process are given in Tables 4, 5,
and 6.  Table 4 contains the pointlike objects in the HDF that are
in color categories 1 and 2.  There is only one candidate in category 1,
ID0094, and it is a lz object.  The remaining entries in Table 4 are
all in category 2.  

Table 5 lists all the pointlike objects in the HDF that we classify
as stars on the basis of having $\Delta < 3\sigma$ and colors in
category 3.  The list contains 16 of the 17 FGB stars and two additional
objects that meet our criteria.  We place the 17$^{th}$ FGB star in Table 4
because it is in our color category 2. As a further confirmation of the
validity of our approach, 5 of our candidate stars (0276, 0341, 1429,
1610 and 3121) have been spectroscopically confirmed by Cohen et
al. (1996) to be indeed stars.

Table 6 lists the 15 compact objects with $\Delta > 3\sigma$, which
we consider to be slightly resolved galaxies. Of these, 10 are in
color category 1, 2 in color category 2, and 3 in color category 3.  
We will consider the objects in the first two categories as potential 
AGN candidates in which the host
galaxy is being detected, although we recognize that
colors alone cannot distinguish them from compact galaxies undergoing
intense star formation.  In fact, during the early development of our
procedure, we identified a compact object with the colors of a
$z \sim 3.5$ quasar.  The object did not remain in our list of candidates
because it was spatially resolved by the procedures we finally adopted for the
selection of compact objects.  However, we do note that Lowenthal et al.
(1996) found it to be an (narrow) emission-line galaxy with $z=3.368$
(our ID0627 = their hdf2\_0705\_1366).
The object is shown as a filled square in Figure 10.  It serves
as both a confirmation of our modeling techniques and an illustration of
the problem of color degeneracy between quasars/AGNs and starbursts. 

\section{Discussion \label{sec:discussion}}
\subsection{Estimate of the Upper Limit on Quasar Candidates in the HDF}

First, we note that there are no candidates with colors indicating
that they have $z > 3.5$; all the candidates in Tables 4 and 6 are 
expected to have $z < 3.5$.  We use the following procedure to estimate 
the upper limit of such quasars and AGNs in the HDF:
\begin{enumerate}

\item There is 1 unresolved object with quasar colors, ID0094.

\item There are 7 unresolved objects whose colors are consistent
with either stars or quasars (the remaining objects in Table 4).

\item There are 10 slightly resolved objects in Table 6 with quasar
colors.  We include them because of the possibility that they
are quasars in which the host galaxy is being detected.

\item There are 2 slightly resolved objects in Table 6 whose colors
are consistent either with stars or quasars.  We include them for
the same reason as given above. 

\end{enumerate}

The sum of the objects listed above is 20, which we take to be our
estimate of the upper limit of quasar and AGN candidates in the HDF  
for $V_{606} \leq 27.0$ and $z < 3.5$.  We surveyed 4.44 arcmin$^2$ of
the HDF\footnote{A 35 pixel border was excluded for each chip to avoid
edge distortions and an incorrect estimate of the sky
background for objects too close to the edge of the chip.} so the
upper limit in surface density is 16,200 per deg$^2$.
Of course, 
spectroscopic observations of the candidates are needed to determine
which objects if any are quasars or AGNs.

How does this upper limit
compare with expected values?  Because the HDF reaches
fainter than any well-studied quasar survey with spectroscopic
confirmation of many objects, one approach is to extrapolate the
results of quantitative surveys done to brighter limiting magnitudes.
For example, the results of Boyle et al. (1988, BSP)
and Warren et al. (1994, WHO) cover the redshift range
$0.1 < z < 4.5$ down to observed magnitudes of about 20
\footnote{The Schmidt et al. (1995, SSG) survey has more
quasars than the WHO survey but is not suitable for
this comparison.  The SSG luminosity function is only valid
for the interval $-27.5 < M_B < -25.5$, where the slope
is steeper than for the less luminous quasars that HDF
can reach.}.
We find that the BSP luminosity function, which is for
$z < 2.2$, predicts 824 quasars per deg$^2$ with $B \leq 27$, 
or 1 quasar in the area of the HDF that we
surveyed, consistent
with our upper limit.  For higher redshift objects, $3.0 < z < 4.5$, 
the WHO luminosity function predicts 2570 quasars 
per deg$^2$ with $R \leq 27$, or 3 quasars in the HDF, also consistent 
with our upper limit.  
Another approach is to use the Hartwick-Schade (1990)
compilation of cumulative surface densities of quasars.  
A linear extrapolation of the log N($<$B) vs. B data in their
Table 3b for $B \geq 19.0$ leads to 8,200 quasars deg$^{-2}$
for $0 < z < 2.2$ and 12,000 quasars deg$^{-2}$ for $0 < z < 3.3$.
These values still within our upper limit although they are
significantly larger than the BSP extrapolation, for example.
However, the BSP luminosity function has a two-power law
form that evolves with redshift.  It is not surprising that
its extrapolation well beyond the limits of the data on which
it is based differs from the simple extrapolation of the
Hartwick-Schade number count data.

The important next step will be to confirm the number of actual
quasars and AGNs in the HDF (for example, with spectroscopic
observations) so that we can pin down the faint end of their luminosity
function at high redshifts.  This is also crucial to estimating the
degree of ionization of the intergalactic medium that
quasars contribute at high redshift (\cite{mir98}). Similarly,
establishing the number of faint quasars with $z<3.5$ is important to
understanding the early formation of quasars and the structures in which
they reside. Haiman et al. (1998) note that the lack of $z>3.5$
candidates in the HDF already is constraining models of structure formation.

\subsection{Faint Blue Galaxies in the HDF}

We concur with other studies of the HDF, e.g., Elson et al. (1996, ESG),
that one of its
striking features is the presence of a significant number of faint,
blue, compact galaxies, many of which are at high redshift and
are regions of intense star formation.  It is likely
that many of the objects in our Table 6 are in this category.  As noted
by ESG, many have nearby companions or show evidence
for having an extended component.  Some are among the bluest objects
in the HDF (Figure 9).  Others are blue in $U_{300}-B_{450}$ but redder
in $B_{450}-V_{606}$ than our model quasars, which is an indication
that they may be compact, narrow, emission-line galaxies. 

\subsection{White Dwarfs in the HDF?}

If white dwarfs in the halo of the Milky Way constitute up to
50\% of the dark matter, as suggested by early MACHO results
(Alcock et al. 1997), then white dwarfs should be observable
in the HDF.  We investigated this possibility with the following
procedure: 1) white dwarf candidates should be unresolved,
$\Delta < 3\sigma$;  2) they should fall in areas of 
the $(\U-\B)/(\B-\V)$ diagram where white
dwarfs would be expected, and 3) any such objects should
also have $(\B-\V)$ and $(\V-\I)$ colors consistent
with white dwarf colors.  We found that objects ID0094, ID0134,
ID0212, and ID1515 meet criteria 1 and 2 but not 3 (they are too red).
Thus, we find no good candidates for white dwarfs in the HDF.
This is either a warning flag for the MACHO results or a signal
that the fraction of white dwarfs brighter than the detectability
limit of the HDF is much smaller than predicted by current halo
white dwarf formation and evolution models (FGB; Kawaler 1996).  In
addition, all the pointlike sources we identify are quite blue and
faint and thus appear too blue to be white dwarfs older than a few Gyrs
and therefore cannot be a significant fraction of the Milky Way halo.

\section{Comparison with Other Searches of the HDF \label{sec:comp}}

Here we compare our results with those of other groups who
have searched the HDF: Williams et al. (1996), FGB,
ESG, and Mendez et al. (1996, MMMBC).
We begin with the Williams et al. article, which provides the
description of the HDF project, the data reduction procedures,
and the catalog of the 3500 objects they found.  Their catalog
gives positions and the main photometric parameters of the objects
and is a fundamental reference for the HDF.  Williams et al.\ used FOCAS,
as we did, to produce their catalog.  They did 
not attempt a detailed analysis of the objects nor a separation
of stars and galaxies.  We found in comparing our list of compact
objects in Table 2 that each object was in the Williams et al.
catalog and that the agreement of the X,Y positions was excellent.

Turning to the FGB work, we have already noted that we recovered
all 17 stars they found.  FGB used the individual dither images
to construct radial profiles for objects and then merged them
to obtain dense sampling, as we have done.  The main difference
is that FGB selected stars by visual inspection of the profiles,
while we used the procedures described in this article.  We found
2 stars in addition to the 17 FGB stars, and 7 more quasar candidates.
This is probably due to our use of the $\Delta$ parameter and
all 4 of the photometric bands in the HDF data.

Next we compare our results with ESG, who used DAOPHOT
to find and make photometric measures of objects in the HDF.  They
list 9 candidate low-mass stars (their Table 1) and 50 faint, blue, 
unresolved objects (their Table 2).  
All 9 star candidates in their Table 1 are in our list of stars, Table 5
\footnote{Note that the X,Y coordinates of both ESG and MMMBC
differ from those of our system and the Williams et al. system.
The offsets are $\Delta X,Y(ESG - Williams) \approx 76,76$.}.
We find in addition 5 objects we consider to be stars, ID0134, ID0273,
ID0341, ID1610, and ID3121.  The 15$^{th}$, a star in our list, ID0276, is
object 2-14 in their Table 2.  Of the 8 quasar candidates in our Table 4,
2 are in their Table 2: ID0563 = their object 2-7 and ID1515 = 3-4.
ESG believed there were no quasar candidates in the HDF, but
their results were based only on measures of the $V_{606}$ and $I_{814}$
images.  We agree with them that there are no high-redshift, $z>3.5$,
candidates in the HDF.  However, as discussed above, we do find
low-redshift candidates from consideration of the $(\U-\B)/(\B-\V)$ diagram.  

Turning to the 50 faint, blue, unresolved objects of ESG,
we find that only 6 of them are in our list of compact objects
in Table 2. There are only 3 other objects in their Table 2 that are
brighter than our classification limit\footnote{Note that our magnitude 
differs from ESG in the sense that  $V_{606}(Conti~et~al.)
- V_{606}(ESG) \approx 0.4$ as determined from the stars measured
in common.  The difference arises from the different apertures and
different photometric routines that were used.}.  
According to ESG,
all 3 are either members of a group or a chain and thus likely
to be parts of galaxies, in accordance with our classification.
The other 41 are fainter than our classification limit.  ESG note
that about half of their faint, blue objects have companions within
1 arcsec and think the objects likely are compact regions of
intense star formation.

Finally, we compare our results with those of MMMBC,
who used SEXTRACTOR on the combined $V_{606}$ and $I_{814}$ drizzled
images to search for stars in the HDF.  They found 14 point-like
objects, of which they believe 8 are stars and 6 are most likely
unresolved extragalactic objects because of their color.  We note
that the 10 brightest objects in their list are all in our list of
stars, Table 5.  The 4 remaining objects are all fainter than
our classification limit.  They considered that there were no
quasar candidates in their lists, but their approach was quite
different from ours.  They did not make use of the $(\U-\B)/(\B-\V)$
color-color plane.  Furthermore, it is not clear to us
how their use of $\V~vs~(\B-\V)$ and $\I~vs~(\V-\I)$ diagrams 
constrains the existence of quasar candidates.  
MMMBC claim a 97\% completeness at $\V \sim 30$, in considerable
disagreement with our determinations, which indicate 50\% completeness
at $V_{606}=29.1$.  

With regard to the 15 compact objects that we list in Table 6 as 
galaxies, we note that none of them were selected by ESG,
FGB, or MMMBC.  Thus there is agreement that these objects
are all slightly resolved.

We conclude from our analysis of the HDF data and
the above comparison of the 4 different investigations:
\begin{enumerate}

\item There is good agreement on the number of stars in the HDF
and on the lack of $z>3.5$ quasar candidates.

\item We find quasar candidates which are expected to have $z<3.5$, in
disagreement with both ESG and MMMBC. We attribute our findings both to
the use of all the available color information and to more detailed
modeling of quasar colors. 

\item Our approach appears to offer better discrimination between
unresolved and resolved objects and is better defined quantitatively
because of our use of the sub-sampled profiles and our modeling of the
detection and classification process. 

\item However, our limiting magnitude for classification
is brighter than those of ESG and MMMBC because of
our use of the individual dithered images for the subsampling. 

\end{enumerate} 
\noindent
Thus, the choice of
approach depends on the goals of the particular investigation.

\section{Future Work: AGNs in the HDF \label{sec:future}}

The present study has concentrated on quasar 
candidates that are starlike or nearly so in the HDF.
This in accord with the classical definition of quasars as starlike
objects of large redshift (\cite{sch70}) and is a logical first step
because starlike objects are the most straightforward to detect and
analyze quantitatively.
However, the classical definition of quasars was developed in the 
era of 1 arcsec 
image quality and limiting magnitudes of order 20.  
The nearly 0.1 arcsec
image quality and 50\% completeness magnitude limits fainter than 28
of the HDF quite literally give us a new image of the universe.  We may
expect two effects to be significant compared to earlier studies: 
1) the host galaxy of a quasar can 
be easier to see, and 2) AGNs of significantly lower luminosity
become detectable at relatively high redshift.  
For example, at $z=2$, an object with observed
magnitude of 27 has an absolute magnitude of $-18$ 
($H_{0}=75,~q_{0}=0.5,~f_{\nu} \propto {\nu}^{-1}$),
five magnitudes fainter than the classical limit for quasars and less 
luminous than many well-known Seyfert galaxies.  Therefore, a search
for resolved objects in the HDF with active galactic nuclei is 
needed.  

What will be good ways to find AGNs in the HDF?  A first step will be to
identify resolved objects whose nuclei have colors consistent with 
AGNs.  However, photometric colors alone will not be sufficient,
because intense starbursts at high redshift
can produce broad-band colors similar to those 
of many quasars, as we have already shown in this paper with our
recovery of the narrow emission-line galaxy at $z=3.368$.  Spectroscopic
observation of broad emission lines and detection of sufficiently powerful
X-ray emission are two of the most promising approaches.  

Spectroscopic results to date indicate that AGNs appear in only about 2\% of
faint galaxies in deep surveys with the Keck telescopes.  Phillips
et al. (1997) found 1 AGN among 54 compact galaxies in the flanking
fields of the HDF.  Steidel reported at the Young Universe meeting
in Monte Porzio in 1997 that 8 of about 350 galaxies with $2.5 < z < 3.5$
and $R \leq 25$ that he and his collaborators had observed were AGNs.
Interestingly, the 2\% figure in combination with the Madau et al.
(1996) results allows us to make another estimate of the number
of AGNs in the HDF.  Madau et al. (1996) used color selection to
identify 69 F300W and 14 F450W dropouts in the HDF. If the 2\% fraction
of AGNs holds for these objects, the expected number of AGNs in
the HDF with $2 < z < 4.5$ is 1.7, comfortably below our estimate
of the upper limit.  Of course spectroscopic
observations are required to establish, for example, what the
number of AGNs is at magnitudes fainter than 25, a range that
has not yet been explored.

Ultra-deep X-ray surveys with 1-2 arcsec accuracy (or better)
positions may offer
a superior approach to this question (\cite{has98}).  
X-ray emission is a strong
signature of activity in galaxies.  The combination of a deep survey
with AXAF, for example, and
deep HST imaging, could provide the most
efficient means of identifying faint AGN candidates for follow-up spectroscopy.
Whatever approach is used, spectroscopic observations will be 
essential to determining 
the relation of quasars and
galaxies at high redshift and to understanding the role AGNs play in the 
formation and evolution of galaxies.

\acknowledgments
We wish to thank A. Gould, L. Davis, H. Ferguson, A. Fruchter and F.
Valdes for useful discussions. We thank the Director of STScI, Robert
Williams, and the HDF team for making available such a rich dataset.
We acknowledge the anonymous referee for comments and suggestions which
helped to improve the overall clarity of the manuscript.
Support for this work was provided by NASA through grant number
AR-07535.01-96A from the
Space Telescope Science Institute, which is operated by Association of
Universities for Research in Astronomy, Incorporated, under NASA
contract NAS5-26555. Support was also provided by NSF grant number
AST-9519324. 

\newpage


\newpage
\begin{deluxetable}{ccccccc}\tablecolumns{7}\tablewidth{0pc}
\tablenum{1}
\footnotesize
\tablecaption{Point Source Completeness for the HDF Crude Combine Images$^{\dag}$ \label{tab:cplt}}
\tablehead{
\colhead{Filter} & \multicolumn{2}{c}{Chip 2} &
\multicolumn{2}{c}{Chip 3} & \multicolumn{2}{c}{Chip 4}\nl
\cline{2-3}
\cline{4-5}
\cline{6-7}
\colhead{} & \colhead{50\%} & \colhead{90\%} & \colhead{50\%} &\colhead{90\%} & \colhead{50\%} & \colhead{90\%}
}
\startdata$U_{300}$ & 26.9 & 26.3 & 26.8 & 26.2 & 26.9 & 26.2 \nl
$B_{450}$ & 28.9 & 28.0 & 28.7 & 28.0 & 28.8 & 28.1 \nl
$V_{606}$ & 29.1 & 27.7 & 29.2 & 28.0 & 29.0 & 27.6 \nl
$I_{814}$ & 28.2 & 26.8 & 28.2 & 26.8 & 28.1 & 26.8 \nl
\enddata
\tablenotetext{\dag}{All magnitudes are in the Vega magnitude system.}
\end{deluxetable}

\begin{deluxetable}{rcccccccccccc}
\tablenum{2}
\tablecolumns{13}
\tablewidth{0pc}
\footnotesize
\tablecaption{Compact Objects in the Hubble Deep Field$^{\dag}$ \label{tab:all_compact}}
\tablehead{
\colhead{ID} & \colhead{Chip} & \multicolumn{1}{c}{RA} &
\multicolumn{1}{c}{Dec} & \colhead{$X$} & 
\colhead{$Y$} & \colhead{$U_{300}$} & \colhead{$B_{450}$} 
& \colhead{$V_{606}$} & \colhead{$I_{814}$} 
& \multicolumn{3}{c}{$\Delta^{\ddag}$} \nl
\cline{11-13}
\colhead{} & \colhead{} & \colhead{(12$^h$)} &
\colhead{(62$^{\circ}$)} & \colhead{} & \colhead{} & 
\colhead{} & \colhead{} & \colhead{} & \colhead{} & \colhead{} &
\colhead{} & \colhead{} \nl
\colhead{} & \colhead{}  & \colhead{m s} &
\colhead{$\prime$ $\prime\prime$} & \colhead{} & \colhead{} & 
\colhead{mag} & \colhead{mag} & \colhead{mag} & \colhead{mag} &
\colhead{F450W} & \colhead{F606W} & \colhead{F814W}
}
\startdata
0094 & 2 & 36 46.40 & 14 08.59 & 1757.76 &  569.62 & 25.99 & 26.98 & 26.82 & 26.40 & 0.23 & 0.28 & 0.34 \nl
0129 & 2 & 36 47.65 & 14 05.78 & 1605.15 &  741.18 & 26.49 & 27.36 & 26.72 & 25.61 & 0.41 & 0.27 & 0.25 \nl
0134 & 2 & 36 47.15 & 14 15.92 & 1873.48 &  763.50 & 24.38 & 25.89 & 25.95 & 25.38 & 0.09 & 0.11 & 0.15 \nl
0212 & 2 & 36 50.14 & 14 10.40 & 1534.37 & 1188.69 & 26.30 & 27.05 & 26.95 & 26.67 & 0.36 & 0.35 & 0.47 \nl
0258 & 2 & 36 51.72 & 14 07.27 & 1350.77 & 1410.35 & 25.64 & 27.09 & 26.98 & 26.50 & 0.31 & 0.34 & 0.38 \nl
0273 & 2 & 36 54.20 & 13 35.87 & 453.45  & 1491.12 &  --   & 26.71 & 26.74 & 26.50 & 0.08 & 0.05 & 0.16 \nl
0276 & 2 & 36 54.73 & 13 28.01 &  235.68 & 1496.09 & 24.18 & 22.07 & 20.43 & 19.37 & 0.00 & 0.02 & 0.33 \nl
0341 & 2 & 36 52.82 & 14 32.05 & 1843.07 & 1836.59 & 23.83 & 22.42 & 21.69 & 20.82 & 0.00 & 0.00 & 0.00 \nl
0357 & 2 & 36 55.53 & 13 58.73 &  885.38 & 1934.88 & 25.51 & 26.99 & 26.92 & 26.75 & 0.18 & 0.23 & 0.53 \nl
0563 & 2 & 36 50.10 & 13 58.20 & 1257.46 & 1057.56 &  --   & 27.41 & 26.81 & 26.30 & 0.23 & 0.07 & 0.29 \nl
0598 & 2 & 36 51.64 & 13 47.40 &  900.36 & 1195.42 &  --   & 27.98 & 26.47 & 25.19 & 0.18 & 0.04 & 0.05 \nl
0661 & 2 & 36 53.12 & 13 46.23 &  768.32 & 1422.07 &  --   & 27.74 & 26.88 & 26.26 & 0.43 & 0.22 & 0.29 \nl
0710 & 2 & 36 54.00 & 13 51.61 &  829.84 & 1617.34 &  --   & 26.84 & 26.98 & 26.51 & 0.12 & 0.10 & 0.27 \nl
1379 & 3 & 36 56.53 & 13 27.41 & 1797.59 &  190.28 & 25.69 & 26.61 & 26.45 & 25.69 & 0.23 & 0.23 & 0.27 \nl
1429 & 3 & 36 53.67 & 13 08.25 & 1144.76 &  431.53 & 25.48 & 23.57 & 22.18 & 21.32 & 0.00 & 0.00 & 0.00 \nl
1477 & 3 & 36 52.97 & 12 56.76 &  918.34 &  647.86 & 25.51 & 26.95 & 26.41 & 25.42 & 0.21 & 0.24 & 0.17 \nl
1495 & 3 & 36 52.25 & 12 49.39 &  729.00 &  767.42 & 25.45 & 26.98 & 26.59 & 25.62 & 0.31 & 0.32 & 0.33 \nl
1515 & 3 & 36 58.56 & 13 05.48 & 1905.62 &  837.85 & 26.35 & 26.57 & 26.39 & 25.86 & 0.13 & 0.15 & 0.17 \nl 
1548 & 3 & 36 54.05 & 12 45.58 &  980.10 &  980.82 & 25.71 & 26.47 & 24.67 & 22.73 & 0.03 & 0.03 & 0.01 \nl
1610 & 3 & 36 56.36 & 12 41.13 & 1306.91 & 1244.63 & 21.48 & 20.36 & 19.94 & 19.29 & 0.10 & 0.33 & 0.36 \nl
1627 & 3 & 36 51.44 & 12 20.74 & 310.88  & 1370.34 & 25.02 & 26.26 & 25.88 & 25.00 & 0.19 & 0.22 & 0.27 \nl 
1670 & 3 & 36 55.63 & 12 23.63 & 1015.46 & 1597.39 &  --   & 26.71 & 26.50 & 26.11 & 0.15 & 0.11 & 0.15 \nl
1721 & 3 & 36 52.60 & 12 01.20 &  303.15 & 1902.13 &  --   & 26.77 & 26.31 & 26.01 & 0.73 & 0.02 & 0.01 \nl
1741 & 3 & 36 53.60 & 13 17.95 & 1231.60 &  203.56 &  --   & 26.57 & 26.28 & 25.94 & 0.19 & 0.18 & 0.22 \nl
1927 & 3 & 36 59.30 & 12 55.80 & 1927.68 & 1112.46 &  --   & 26.47 & 24.90 & 22.54 & 0.03 & 0.01 & 0.07 \nl
2086 & 3 & 36 55.38 & 12 13.40 &  872.46 & 1815.47 &  --   & 28.46 & 26.93 & 24.50 & 0.77 & 0.02 & 0.01 \nl
2103 & 3 & 36 52.56 & 12 01.70 &  301.32 & 1887.65 &  --   & 26.02 & 24.60 & 23.72 & 0.02 & 0.08 & 0.00 \nl
3027 & 4 & 36 48.62 & 12 07.80 & 1469.35 &  510.24 & 25.89 & 27.12 & 26.78 & 25.95 & 0.43 & 0.27 & 0.39 \nl
3112 & 4 & 36 46.58 & 11 57.16 & 1575.97 &  944.23 & 25.81 & 26.94 & 26.18 & 25.73 & 0.34 & 0.40 & 0.53 \nl
3121 & 4 & 36 45.41 & 12 13.57 & 1116.71 &  971.88 & 20.32 & 21.16 & 21.22 & 20.86 & 0.00 & 0.00 & 0.00 \nl
3134 & 4 & 36 43.79 & 12 32.09 &  577.30 & 1053.60 & 25.07 & 26.62 & 26.53 & 26.12 & 0.31 & 0.33 & 0.39 \nl
3148 & 4 & 36 45.78 & 11 50.55 & 1674.50 & 1136.57 & 26.12 & 26.57 & 26.14 & 25.73 & 0.26 & 0.30 & 0.33 \nl
3176 & 4 & 36 44.73 & 11 57.07 & 1451.73 & 1243.17 & 25.04 & 26.37 & 25.78 & 24.94 & 0.12 & 0.29 & 0.28 \nl
3177 & 4 & 36 42.86 & 12 27.87 &  611.79 & 1244.15 & 25.91 & 26.23 & 26.10 & 25.67 & 0.22 & 0.22 & 0.26 \nl
3221 & 4 & 36 44.60 & 11 39.45 & 1849.88 & 1437.03 &  --   & 26.86 & 26.66 & 26.31 & 0.23 & 0.29 & 0.20 \nl
3226 & 4 & 36 44.40 & 11 39.05 & 1846.01 & 1472.08 & 25.40 & 26.99 & 26.82 & 26.34 & 0.34 & 0.35 & 0.33 \nl
3227 & 4 & 36 42.90 & 12 03.46 & 1178.63 & 1477.11 &  --   & 27.11 & 26.88 & 26.55 & 0.34 & 0.33 & 0.45 \nl
3418 & 4 & 36 45.98 & 12 50.38 &  304.23 &  521.25 &  --   & 26.63 & 25.36 & 24.23 & 0.05 & 0.04 & 0.05 \nl
3421 & 4 & 36 46.76 & 12 37.06 &  666.06 &  524.04 &  --   & 25.86 & 24.14 & 22.32 & 0.02 & 0.05 & 0.04 \nl
3541 & 4 & 36 45.41 & 11 48.95 & 1685.95 & 1213.05 &  --   & 27.17 & 26.62 & 26.18 & 0.36 & 0.34 & 0.34 \nl
3595 & 4 & 36 40.88 & 12 34.00 &  334.51 & 1504.03 &  --   & 27.51 & 25.71 & 23.98 & 0.19 & 0.06 & 0.02 \nl
\enddata
\tablenotetext{\dag}{Celestial coordinates are in epoch J2000. 
We used chip coordinates according to Williams et al. (1996).}
\tablenotetext{\ddag}{This parameter represents the degree of deviation
from the stellar profile of all the objects in the HDF selected as
compact based on their measured $I_0$ value. Smaller numbers
correspond to smaller deviations (see Section~4). The
distribution of expected deviations given by our simulations is shown
in Table~3.} 
\end{deluxetable}

\begin{deluxetable}{ccccccc}
\tablenum{3}
\tablecolumns{7}\tablewidth{0pc}
\footnotesize\tablecaption{Variation of $\Delta$ as a function of magnitude$^{\dag}$
\label{tab:delta}}\tablehead{
\colhead{Magnitude} & \multicolumn{2}{c}{$\Delta_{B_{450}}$} &\multicolumn{2}{c}{$\Delta_{V_{606}}$} & \multicolumn{2}{c}{$\Delta_{I_{814}}$}\nl
\cline{2-3}\cline{4-5}
\cline{6-7}\colhead{} & \colhead{$\sigma$} & \colhead{$3\sigma$} & \colhead{$\sigma$} &
\colhead{$3\sigma$} & \colhead{$\sigma$} & \colhead{$3\sigma$}}
\startdata23 & 0.022 & 0.066 & 0.021 & 0.063 & 0.011 & 0.034 \nl
24 & 0.021 & 0.064 & 0.022 & 0.065 & 0.018 & 0.053 \nl25 & 0.035 & 0.104 & 0.023 & 0.068 & 0.020 & 0.060 \nl
26 & 0.055 & 0.166 & 0.035 & 0.106 & 0.058 & 0.174 \nl27 & 0.090 & 0.271 & 0.076 & 0.228 & 0.255 & 0.764 \nl
28 & 0.428 & 1.285 & 0.309 & 0.927 & 0.440 & 1.321 \nl
\enddata
\tablenotetext{\dag}{Defined in Section~4 as a measure of the departure of an
object from the HDF PSF in the appropriate filter, $\Delta$ has been
measured here for our simulated stars in order to provide a range of
allowed variation as a function of magnitude. This range will be
used to classify the 41 compact objects in the HDF.}
\end{deluxetable}

\begin{deluxetable}{rccccc}
\tablenum{4}
\tablecolumns{11}
\tablewidth{0pc}
\footnotesize
\tablecaption{Quasar Candidates in the Hubble Deep Field \label{tab:cand}}
\tablehead{
\colhead{ID} & \colhead{Chip} & \colhead{$U_{300} - B_{450}$} & \colhead{$B_{450} - V_{606}$} &\colhead{$V_{606} - I_{814}$} & \colhead{Notes$^{\dag}$} \nl
}\startdata
0094 & 2 & -0.99 & 0.16 & 0.43 & q (lz) \nl0212 & 2 & -0.75 & 0.10 & 0.28 & qs (lz) \nl
0258 & 2 & -1.45 & 0.11 & 0.48 & qs (lz) \nl0563 & 2 &   --  & 0.60 & 0.51 & qsf (hz) \nl
1515 & 3 & -0.22 & 0.18 & 0.53 & qs (lz) \nl1670 & 3 &   --  & 0.21 & 0.39 & qs (hz) \nl
1721 & 3 &   --  & 0.46 & 0.30 & qs (hz) \nl1741 & 3 &   --  & 0.30 & 0.34 & qs (hz) \nl
\enddata\tablenotetext{\dag}{All objects have
$\Delta<3\sigma$. Objects denoted by {\it q} have colors within the quasar 
locus and away from the stellar locus; objects denoted by {\it qs} have 
colors consistent with either quasars or stars. The designation {\it lz} is 
for objects which are UV bright, fall in the quasar locus in Figure~9, and 
are expected to have $z<3$. Similarly, {\it hz} is for objects in the quasar 
locus in Figure~10 expected to have $z>3$. However, for most of these objects 
there is a significant overlap with the stellar locus.}
\end{deluxetable}

\begin{deluxetable}{rccccc}
\tablenum{5}
\tablecolumns{6}
\tablewidth{0pc}
\footnotesize
\tablecaption{Stars in the Hubble Deep Field \label{tab:stars}}
\tablehead{
\colhead{ID} & \colhead{Chip} & \colhead{$U_{300} - B_{450}$} & \colhead{$B_{450} - V_{606}$} &
\colhead{$V_{606} - I_{814}$} & \colhead{Notes$^{\dag}$}
}
\startdata
0134$\phm{a}$ & 2 & -1.50 & -0.06 & 0.57 & s   \nl
0273$\phm{a}$ & 2 &   --  & -0.03 & 0.25 & s   \nl
0276$^a$      & 2 &  2.12 &  1.65 & 1.06 & sf  \nl
0341$^a$      & 2 &  1.42 &  0.72 & 0.87 & sf  \nl
0357$\phm{a}$ & 2 & -1.49 &  0.07 & 0.17 & scf \nl
0598$\phm{a}$ & 2 &   --  &  1.52 & 1.28 & scf \nl
0710$\phm{a}$ & 2 &   --  & -0.14 & 0.47 & scf \nl
1429$^a$      & 3 &  1.91 &  1.39 & 0.86 & sf  \nl
1548$\phm{a}$ & 3 & -0.76 &  1.80 & 1.92 & sf  \nl
1610$^a$      & 3 &  1.12 &  0.43 & 0.65 & sf  \nl
1927$\phm{a}$ & 3 &   --  &  1.57 & 2.37 & sf  \nl
2086$\phm{a}$ & 3 &   --  &  1.53 & 2.43 & scf \nl
2103$\phm{a}$ & 3 &   --  &  1.41 & 0.89 & sf  \nl
3121$^a$    & 4 & -0.85 & -0.06 & 0.36 & sf  \nl
3221$\phm{a}$ & 4 &   --  &  0.20 & 0.35 & scf \nl
3418$\phm{a}$ & 4 &   --  &  1.27 & 1.12 & sf  \nl
3421$\phm{a}$ & 4 &   --  &  1.71 & 1.83 & sf  \nl
3595$\phm{a}$ & 4 &   --  &  1.80 & 1.73 & sf  \nl
\enddata
\tablenotetext{\dag}{Labeled as {\it s} are all the objects we haveclassified as stars in the HDF. Objects classified as stars
by Flynn et al. (1996) are labeled as {\it f}. Objects denoted by {\it c} are stars from the FGB sample that have been re-examined.}
\tablenotetext{a}{Spectroscopically confirmed star. See Cohen et
al. (1996) for details.}
\end{deluxetable}

\begin{deluxetable}{rccccc}
\tablenum{6}
\tablecolumns{6}
\tablewidth{0pc}
\footnotesize
\tablecaption{Compact Galaxies in the Hubble Deep Field\label{tab:galaxies}}
\tablehead{
\colhead{ID} & \colhead{Chip} & \colhead{$U_{300} - B_{450}$} & \colhead{$B_{450} - V_{606}$} &
\colhead{$V_{606} - I_{814}$} & \colhead{Notes$^{\dag}$}
}
\startdata
0129 & 2 & -0.87 & 0.64 & 1.12 & g \nl
0661 & 2 &   --  & 0.86 & 0.62 & qs (hz)\nl1379 & 3 & -0.92 & 0.16 & 0.76 & q (lz)\nl
1477 & 3 & -1.44 & 0.54 & 1.00 & q (lz)\nl1495 & 3 & -1.53 & 0.40 & 0.96 & q (lz)\nl
1627 & 3 & -1.24 & 0.38 & 0.88 & q (lz)\nl3027 & 4 & -1.23 & 0.34 & 0.83 & q (lz) \nl
3112 & 4 & -1.13 & 0.77 & 0.45 & g \nl3134 & 4 & -1.55 & 0.09 & 0.41 & q (lz) \nl
3148 & 4 & -0.45 & 0.43 & 0.41 & q (lz)$^{\ddag}$ \nl3176 & 4 & -1.33 & 0.59 & 0.84 & q (lz) \nl
3177 & 4 & -0.32 & 0.13 & 0.42 & q (lz) \nl3226 & 4 & -1.58 & 0.17 & 0.48 & q (lz) \nl
3227 & 4 &   --  & 0.23 & 0.33 & qs (hz) \nl3541 & 4 &   --  & 0.55 & 0.44 & g  \nl
\enddata\tablenotetext{\dag}{All these objects are compact galaxies for which
$\Delta>3\sigma$. Objects denoted by {\it q} have colorswithin the
quasar locus and away from the stellar locus; objects denoted 
by {\it qs} have colors consistent with either quasars or stars;
objectsdenoted by {\it g} have colors not consistent with quasars. The 
designation {\it lz} and {\it hz} have the same meaning as in Table~4.}
\tablenotetext{\ddag}{Falls within the locus of low $z$ quasars in
Figure~9 but is on the stellar locus in Figure~10.}
\end{deluxetable}

\begin{figure}
\plotfiddle{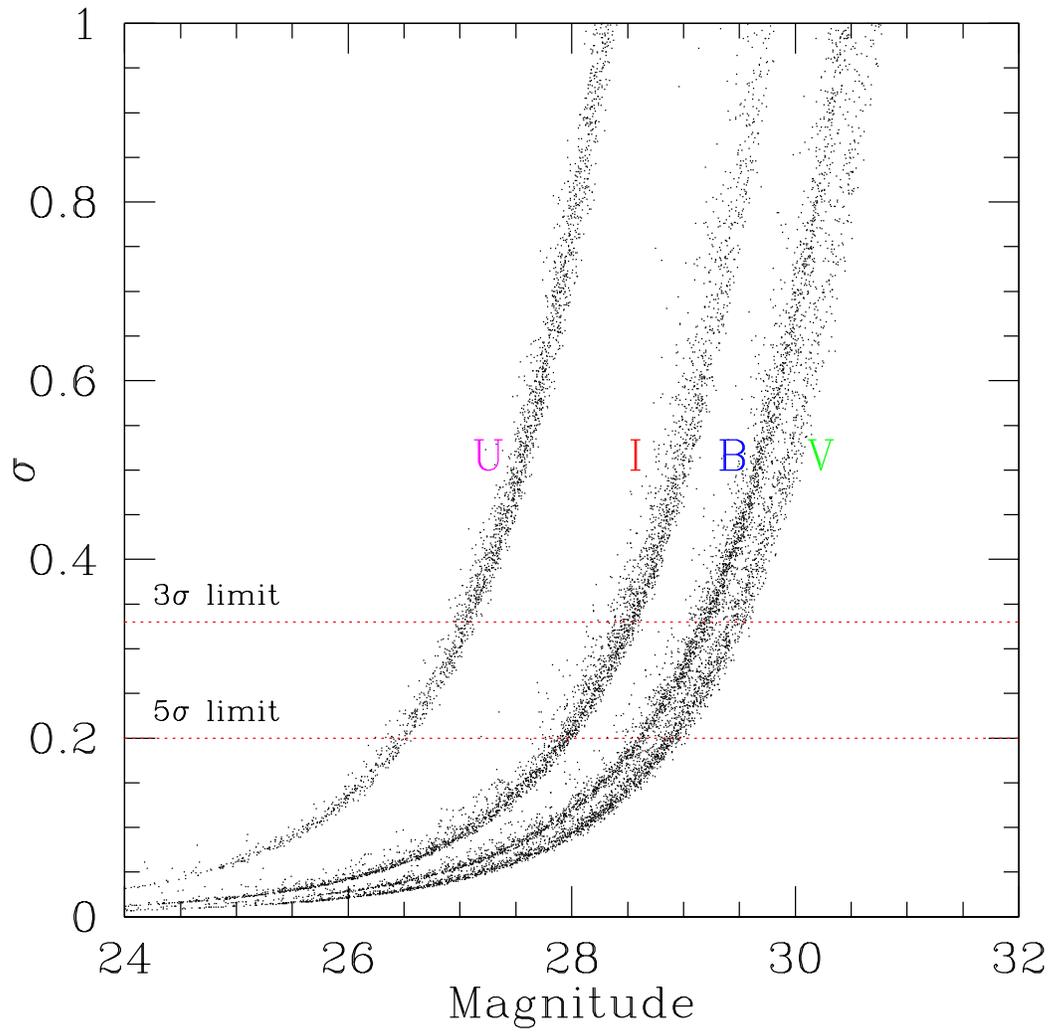}{5.0truein}{0}{70}{70}{-220}{-40}
\caption{Photometric error as a function of magnitude in
all the HDF filters as computed by IRAS/PHOT for all the objects 
in the Hubble Deep Field. The dotted lines represent the 3 and 5~
$\sigma$ magnitude limits. \label{fig:magerr}} 
\end{figure}

\begin{figure}
\plotfiddle{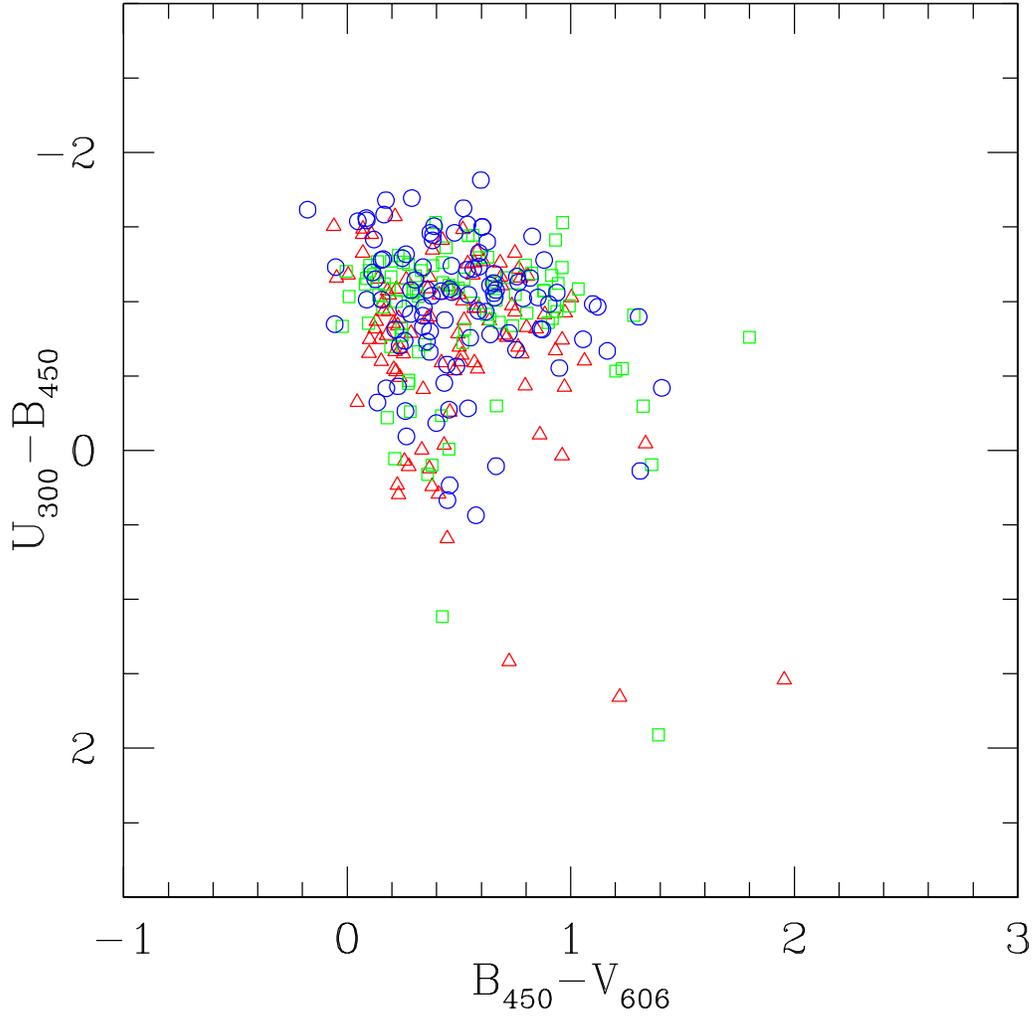}{5.0truein}{0}{70}{70}{-220}{-50}
\caption{$\UBBV$ color-color diagram of all the objects in
the HDF with a $5\sigma$ detection in all four bands.
Open triangles represent sources in Chip 2,
squares in Chip 3 and circles in Chip 4 respectively. The lack of a 
red population of objects is evident. Most of the objects have
$-1.8<\UB<0$ and $0<\BV<1$. From our detailed analysis of the
morphology of the bluest objects, we concluded that the blue sources
that dominate the counts are likely to be regions of intense star
formation rather than be of stellar origin. \label{fig:ubbv}}
\end{figure}

\begin{figure}
\plotfiddle{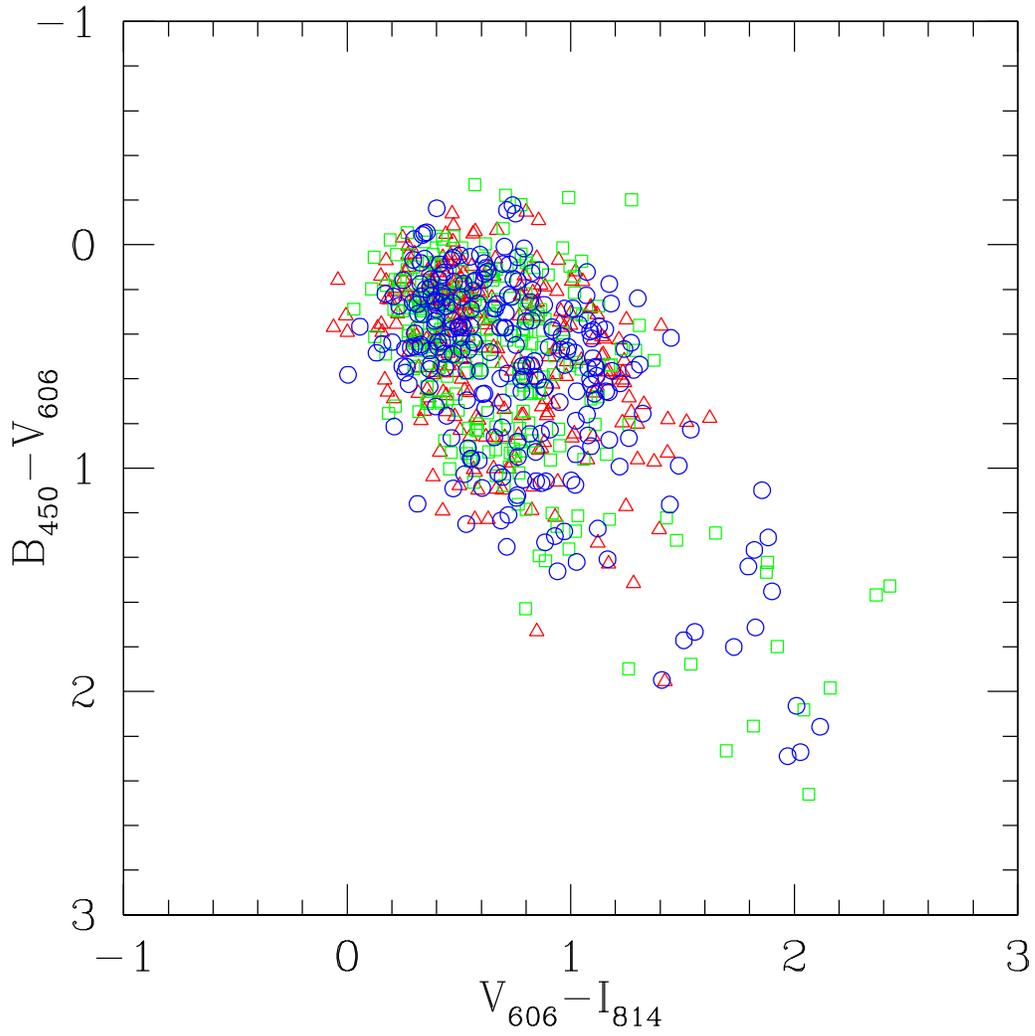}{5.0truein}{0}{70}{70}{-220}{-50}
\caption{$\BVVI$ color-color diagram of all the objects in
the HDF with a $5\sigma$ detection in all four bands. 
Open triangles represent sources in Chip 2,
squares in Chip 3 and circles in Chip 4. The dominance of
the blue population of objects is evident. The bulk of the objects
have $-0.1<\BV<1.2$ and $0.2<\VI<1.3$. Our morphological analysis
excludes a stellar origin for the great majority of this blue
population. \label{fig:bvvi}}
\end{figure}

\begin{figure}
\plotfiddle{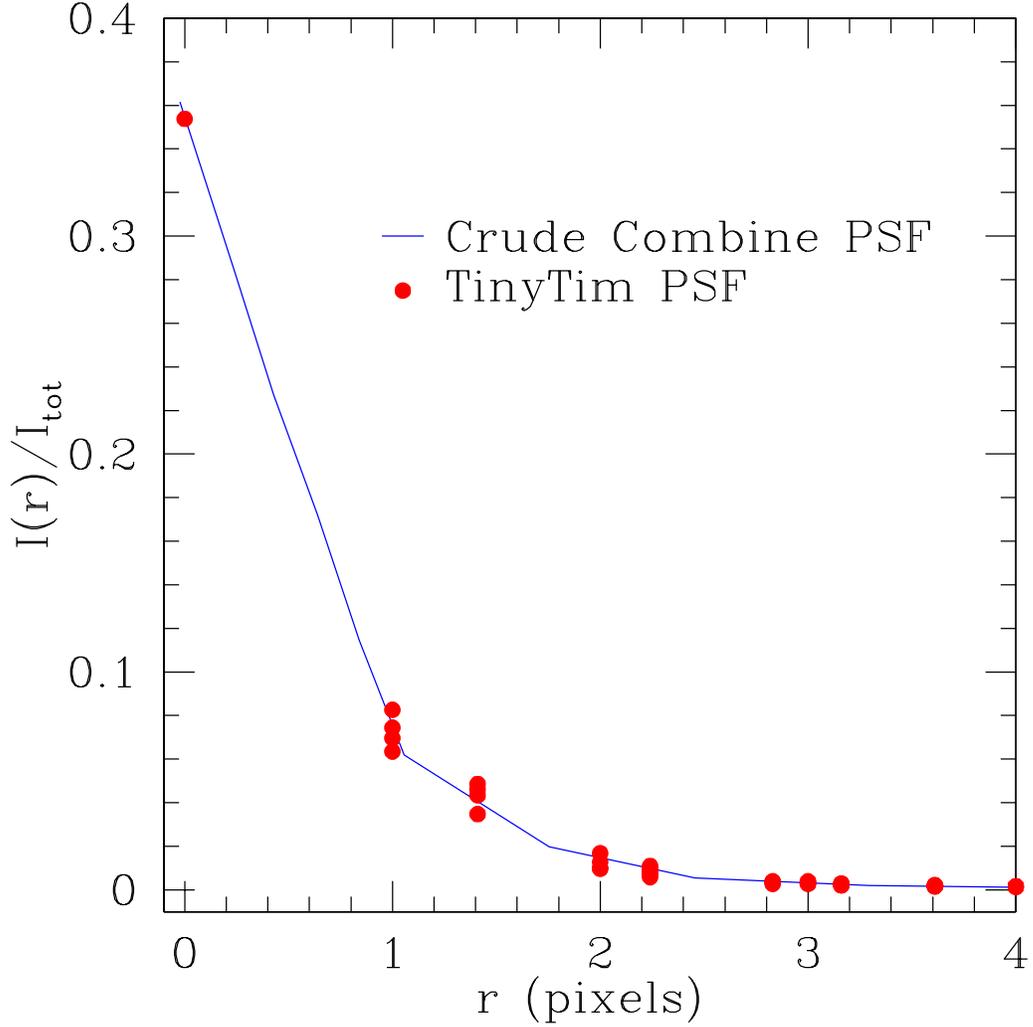}{5.0truein}{0}{70}{70}{-220}{-70}
\caption{The crude combine PSF template used for
Chip 4 in the 
$\I$ filter. Shown is the distance in pixels from the
center of the template against the fraction of light falling on that
pixel. The flux has been normalized to the total flux within a 4 pixel
radius aperture. The WFPC2 undersampling is evident. The filled circles
represent the PSF in the $\I$ filter generated by TinyTim that was
used as a comparison to that of the actual crude combine images. As clearly
shown, TinyTim implicitly makes the assumption that the source is
always centered on the central pixel. Real
sources will be randomly positioned with respect to the central pixel
and hence produce a PSF that samples a wider range of distances
from the center. A separate template for each chip and filter was used
to determine the completeness of the dataset to point
sources. \label{fig:psf-crude}}
\end{figure}

\begin{figure}
\plotfiddle{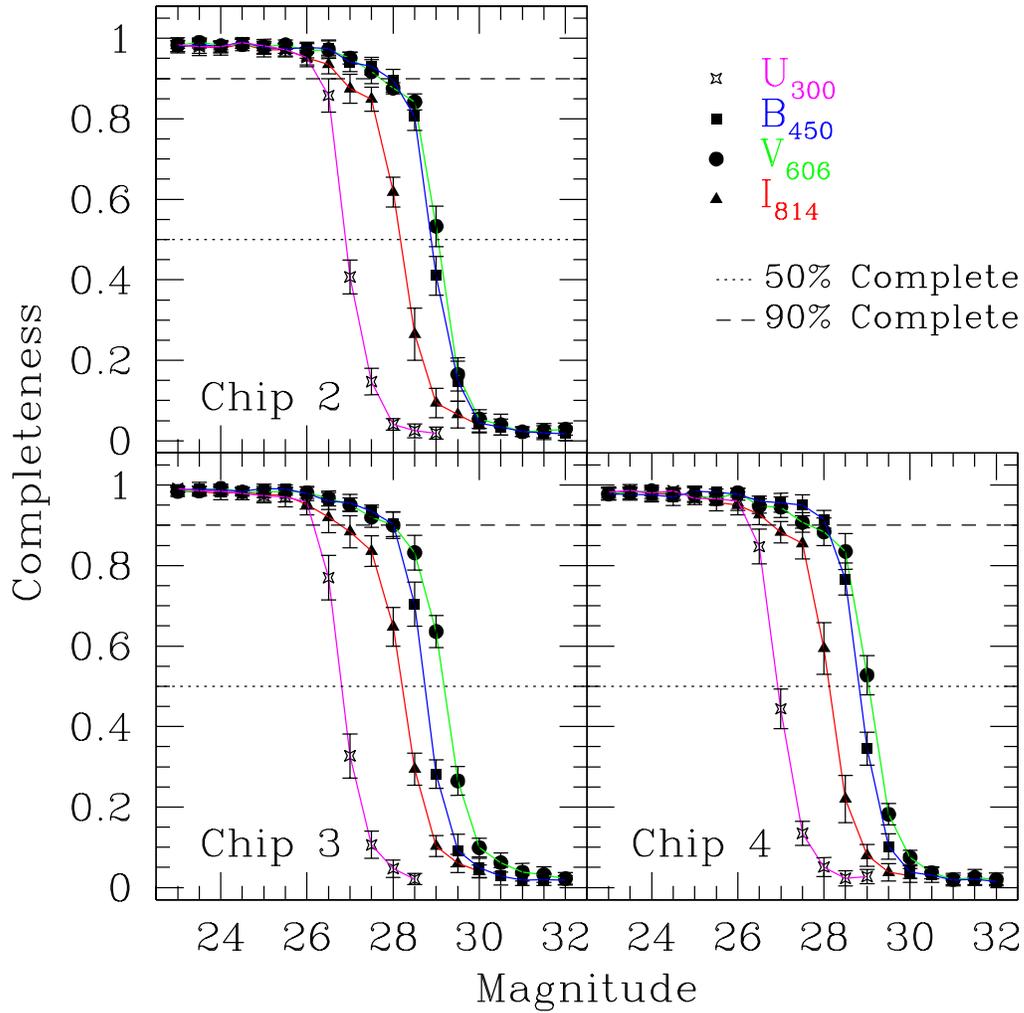}{5.0truein}{0}{70}{70}{-220}{-60}
\caption{Completeness estimates for point sources in the Crude
Combine images 
of the Hubble Deep Field. Shown is the fraction of added template
point sources that were recovered as a function of magnitude.
The simulations were run 10 times each and the error bars 
represent a 1$\sigma$ spread around the mean value.
Notice the slow linear decline before the sharp cutoff at the 
detection threshold. This change in behavior occurs at the
$\sim 90\%$ completeness level. The plots are arranged as the actual
WFPC2 chips on board Hubble, Chip 2 being on the top left corner,
Chip 3 on the bottom left and Chip 4 on the bottom right. \label{fig:cplt}}
\end{figure}

\begin{figure}
\plotfiddle{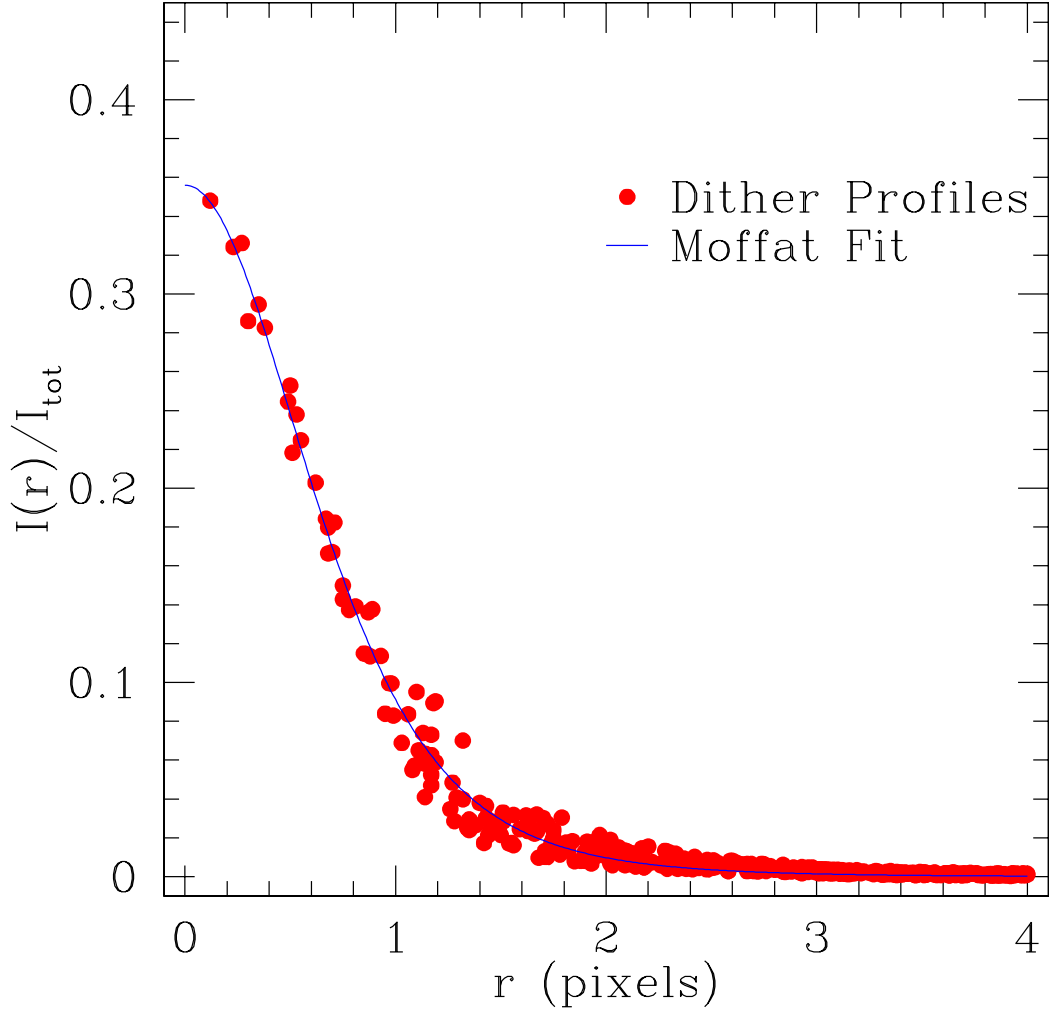}{5.0truein}{0}{70}{70}{-220}{-70}
\caption{For classification purposes we need to be
able to locate 
the center of the object with an accuracy of better than 0.1 pixels.
This is accomplished by making use of the symmetry of the image about
its center (see \cite{fly96}). By using single dither positions,
we can effectively sample the objects' radial profile well within the
inner pixel, thereby overcoming the WFPC2 undersampling. The filled
circles represent the 11 different $\V$ band dithers for a star in Chip
2. The solid line is a Moffat fit to the profile and it will be used 
as a ``functional approximation'' to the radial profile. This will
allow for a systematic approach to object
classification. \label{fig:psf-dither}}
\end{figure}

\begin{figure}
\plotfiddle{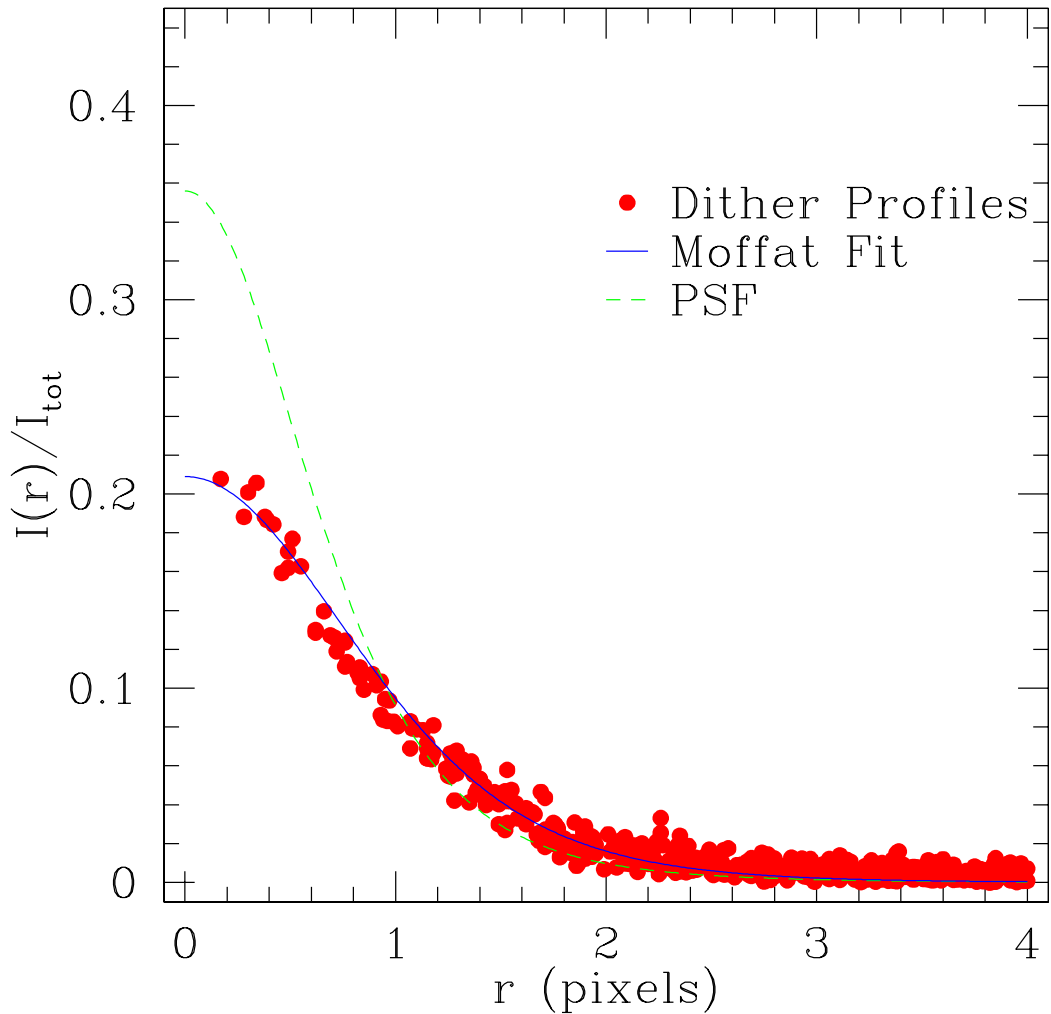}{5.0truein}{0}{70}{70}{-220}{-50}
\caption{$\V$ band radial profile of a galaxy in Chip 4. The
combined profile (filled circles) was obtained by adding together all
the different radial profiles obtained from the 11 dither exposures.
The solid line is a Moffat fit to the combined profile and the PSF is
also shown for comparison. The central intensity parameter of the
Moffat fit $I_0$ is indeed well suited for use as a star/galaxy
separator. \label{fig:galaxy}}
\end{figure}

\begin{figure}
\plotfiddle{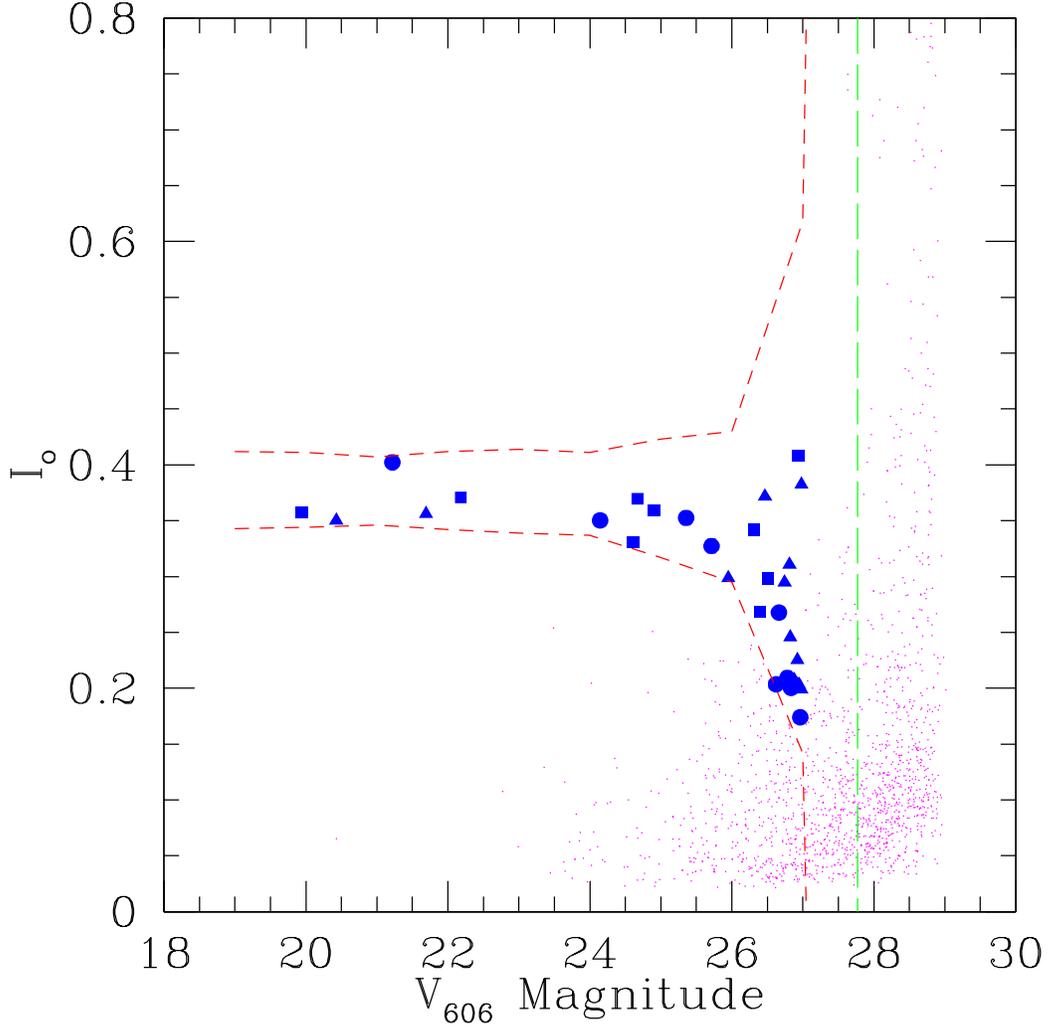}{5.0truein}{0}{70}{70}{-220}{-60}
\caption{$\V$ stellar candidates in the HDF. The central
intensity parameter $I_0$ is shown as a function of magnitude for all
the objects in the HDF. Triangles represent sources in Chip 2,
squares in Chip 3 and circles in Chip 4.
Candidates are selected based on the expected variation of $I_0$ with
magnitude given by our simulations. The area within the dotted lines represents the
$95\%$ confidence interval within which simulated stellar objects lie. As we
approach the completeness limits  of the dataset ($\V=27.8$, shown as
a dashed line), $I_0$ becomes progressively unable to classify objects
reliably. \label{fig:Vclass}}
\end{figure}

\begin{figure}
\plotfiddle{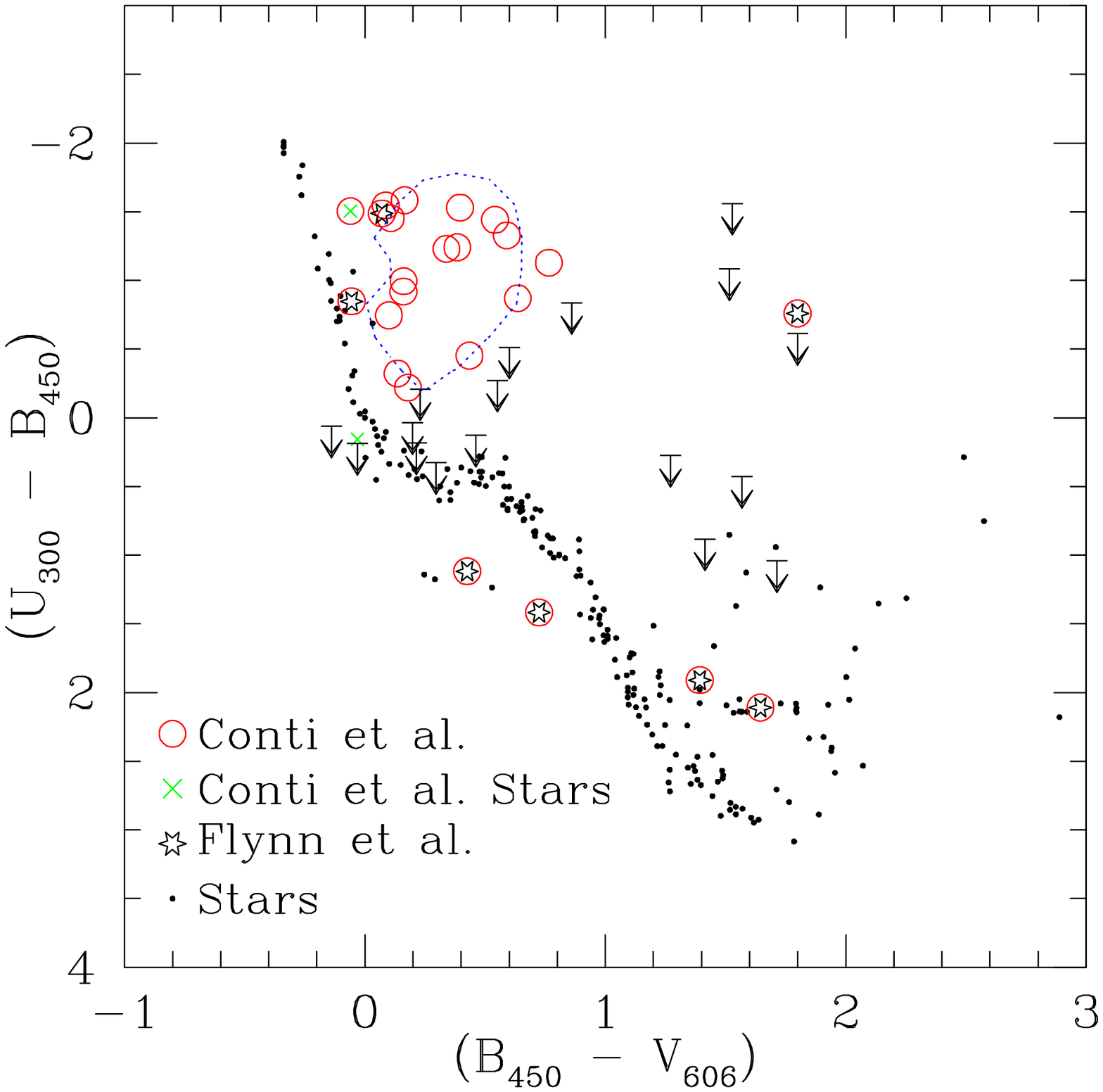}{5.0truein}{0}{70}{70}{-220}{-60}
\caption[lowz.con.ps]{$\UBBV$ color-color diagram of the most compact
sources in the HDF (shown as open circles). Objects classified as
stars by \cite{fly96} are shown as open stars, while new star candidates
are shown as crosses. Filled circles
represent stars from the Bruzual-Persson-Gunn-Stryker spectrophotometric
atlas and are plotted to outline the stellar locus. The contour outlines
our simulated low redshifts ($z<2.2$) quasar colors. The contour encloses 
99\% our models. Limiting colors are also shown for those objects that do
not have a $\U$ detection. Table~4 lists the main
properties of our quasar candidates.\label{fig:lowz}}
\end{figure}

\begin{figure}
\plotfiddle{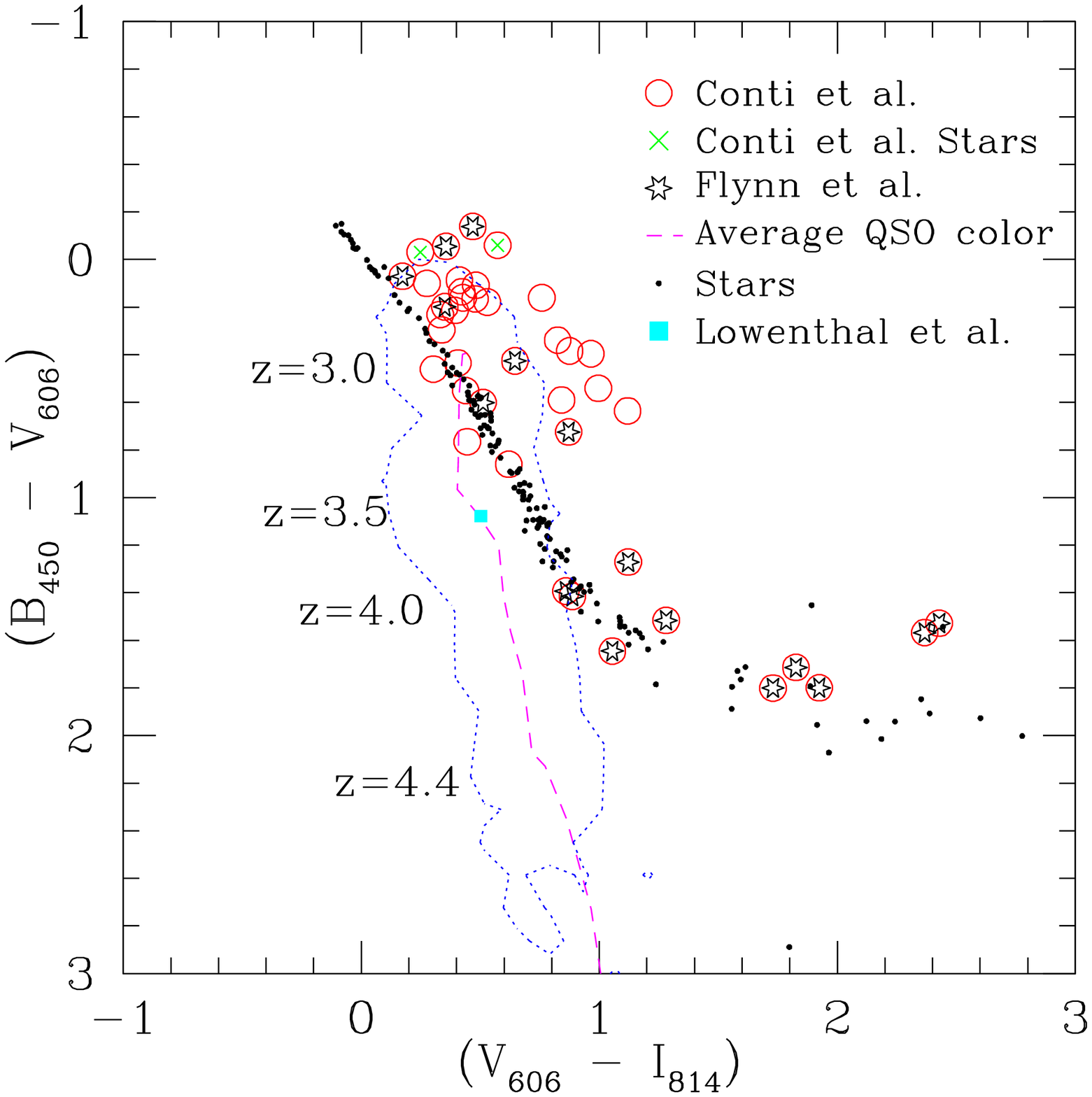}{5.0truein}{0}{70}{70}{-220}{-60}
\caption[highz.con.ps]{$\BVVI$ color-color diagram of the most compact
sources in the HDF (shown as open circles). Objects classified as
stars by \cite{fly96} are shown as open stars, while new star candidates
are shown as crosses.
Filled circles represent stars from the
Bruzual-Persson-Gunn-Stryker spectrophotometric 
atlas and are plotted to outline the stellar locus. 
The contour contains 99\% of our high redshifts ($z>3$) model quasars.
The dashed line shows the average quasars color as a function of redshift.
Also shown as a filled square is a known
emission-line galaxy with $z=3.368$ (\cite{low96}) which was
independently recovered as a relatively bright compact
object during the early development stage of the
classification algorithm. \label{fig:highz}}
\end{figure}

\end{document}